\documentclass[preprint]{aastex}
\usepackage{psfig}
\newcommand{\myv}{V$_{606}$}
\newcommand{\myi}{I$_{814}$}
\newcommand{\mystis}{I$_{\rm LP}$}
\newcommand{\mycolor}{V$_{606}$--I$_{814}$}

\slugcomment{to appear in the January 2001 issue of the PASP}

\begin{document}

\title{HST Color-Magnitude Data for Globular Clusters: I. Transformations Between STIS LP Magnitudes and WFPC2 F606W and F814W Magnitudes\footnotemark[1]}

\author{Mark L.~Houdashelt, Rosemary F.~G.~Wyse}
\affil{Department of Physics \& Astronomy, Johns Hopkins University, \\ 
3400 North Charles Street, Baltimore, MD 21218}

\and

\author{Gerard  Gilmore}
\affil{Institute of Astronomy, University of Cambridge, \\
Madingley Road, Cambridge CB3 0HA, UK}

\footnotetext[1]{Based on observations with the NASA/ESA {\it Hubble Space Telescope}, obtained at the Space Telescope Science Institute, which is operated by AURA, Inc., under NASA contract NAS5-26555.}

\begin{abstract}

We present deep {\it Hubble Space Telescope} optical imaging
observations of two Galactic globular clusters, 47~Tuc and M15, using
the {\it Space Telescope Imaging Spectrograph} (STIS) longpass (LP) filter
and the {\it Wide Field and Planetary Camera 2} (WFPC2)
F606W and F814W filters.  These globular clusters have very
different metallicities (${\rm [Fe/H] \sim -0.7}$~dex for 47~Tuc, and
${\rm [Fe/H] \sim -2.2}$~dex for M15), essentially spanning the
metallicity distribution of the entire Milky Way globular cluster
system, and were chosen to investigate the relationship between
magnitudes in the non-standard STIS LP system and the better
characterized WFPC2 magnitudes.  We determine this relationship by
combining our new STIS data with archival WFPC2 data, taking care to
provide a robust and reliable transformation.  Examining the 47 Tuc
and M15 data separately, there is no evidence for a metallicity
dependence in the transformation relations.  This allows us to combine
the data for both clusters, and also include the stars in the Small
Magellanic Cloud that lie within the field of view of the 47~Tuc images,
to determine final transformations that are valid for main sequence
stars with colors in the range $0.3 \lesssim$ \mycolor\ $\lesssim 1.8$. 
These final relations predict STIS
magnitudes within 0.05 mag (1$\sigma$) of those measured for stars in both
clusters.  In an appendix, we discuss the differences between our empirical
transformation relations and those predicted by synthetic photometry.

\end{abstract}


\section{Introduction}

We have obtained deep {\it Hubble Space Telescope} (HST) observations of the
outer regions of two Galactic globular clusters, 47~Tuc (NGC~104) and M15
(NGC~7078), using the {\it Space Telescope Imaging Spectrograph}
(STIS) and the {\it Near-Infrared Camera and Multi-Object
Spectrometer} (NICMOS), as supporting observations for a project
(GO~7419) with the scientific goal of determining the faint stellar
luminosity function (LF) in the apparently dark-matter-dominated Ursa Minor
dwarf spheroidal galaxy (UMi~dSph).  These data are supplemented by
archival {\it Wide-Field and Planetary Camera 2} (WFPC2) observations
of these clusters and were obtained to aid our understanding of the deep
STIS data for the Ursa Minor galaxy.  The UMi~dSph is one of the
few Galactic satellites which appears to have undergone a very
short-lived episode of star formation; indeed, its stellar population
is remarkably similar to that of a classical halo globular cluster
such as M15, being old and metal-poor \citep{mateo}.  Thus, a direct
comparison between the luminosity functions of M15
and the UMi dSph is equivalent to a stellar mass function comparison
and constrains the amount of baryonic dark matter in the UMi dSph
(see Feltzing, Gilmore \& Wyse 1999 and Wyse et al.~1999 for preliminary
analyses that find remarkable similarity between the LFs of the UMi~dSph,
M15 and M92, another metal-poor halo globular cluster).

In this paper, we discuss the application of our multi-instrument
photometry of 47~Tuc and M15 to the derivation of transformations from
one photometric system to another.  Here, we use aperture photometry of
stars in these clusters to derive transformations between the WFPC2
data taken with the F606W and F814W filters and STIS photometry taken
with the optical longpass (LP) filter, including an investigation of
the possible existence of any metallicity dependence introduced by the
broad STIS LP filter.  Unless otherwise noted, we specify the sizes of all
photometric apertures in terms of their radii (not their diameters)
throughout this paper.  In Section~2, we describe the observational
data and their reduction.  Section~3 discusses the photometry and how
we selected the stars used in calculating the photometric
transformations.  Section~4 presents the resulting transformation
equations, and further discussion of these results is given in
Section~5.  We summarize our conclusions in Section~6, and Appendix~A describes
the differences between our empirical transformation relations and the relations
predicted by SYNPHOT.
In a complementary paper, \citet{beaulieu} derive similar calibrations
from STIS LP and WFPC2 F555W and F814W observations of NGC~6553, a
metal-rich (${\rm [Fe/H] \sim -0.2}$) and highly reddened
(E$_{\rm B-V}\sim 0.7$) globular cluster.

\section{Observations and Data Reduction}

For both 47~Tuc and M15, the pointing and orientation of HST were
chosen so that the fields observed with STIS and NICMOS
would lie within regions of these clusters for which archive WFPC2 images were
available.  Here, we discuss only the STIS and WFPC2 data; the NICMOS
observations will be presented in a future paper.
Table~\ref{hstdata} summarizes the observational parameters of the HST
datasets that we used to obtain WFPC2 and STIS magnitudes of stars in
47~Tuc and M15.  All of the data reduction was performed using standard STSDAS
tasks within IRAF\footnotemark[2].

\footnotetext[2]{IRAF is distributed by National Optical Astronomy Observatories, operated by the Association of Universities for Research in Astronomy, Inc., under contract with the National Science Foundation.}

\subsection{STIS}

The STIS field for M15 was centered at $\alpha_{2000}$~=~21$^{\rm h}$~29$^{\rm m}$~42.$^{\rm s}$43,
$\delta_{2000}$~=~12$\arcdeg$~11$\arcmin$~18.$^{\prime\prime}$6, a region
4.$^{\prime}$6 ($\sim$65 r$_{\rm c}$; Harris 1996) from the cluster's
center that had been
observed with WFPC2 as part of a GTO parallel program (HST proposal 5092).
For 47~Tuc, STIS was also centered 4.$^{\prime}$6 ($\sim$10 r$_{\rm c}$; Harris 1996) from the cluster's center,
at $\alpha_{2000}$~=~0$^{\rm h}$~25$^{\rm m}$~17.$^{\rm s}$26,
$\delta_{2000}$~=~--72$\arcdeg$~5$\arcmin$~36.$^{\prime\prime}$9, in a field which had been
observed with WFPC2 as part of another GTO parallel program (HST proposal 5091)
and also as part of the GO HST Medium-Deep Survey (HST proposal 5369).

The characteristics of STIS and the on-orbit performance of this
instrument are described by \citet{kimble}.  For all of the STIS
observations of 47~Tuc and M15 discussed here, we used gain=4,
CR-SPLIT=5, the STIS optical CCD, which consists of 1024~$\times$~1024
pixels and has a plate scale of 0.$^{\prime\prime}$05071 pixel$^{-1}$,
and the optical longpass (LP) filter, which provides a rather flat
throughput longwards of $\sim6000$ \AA\ to the 1$\mu$m limit (and
reduces the STIS effective field of view to approximately
28\arcsec~$\times$~51\arcsec).  We will refer to magnitudes measured
through this filter as \mystis\ magnitudes; these are non-standard
magnitudes that we relate here to the more conventional WFPC2 systems. 

Data reduction for the STIS observations of both clusters involved a simple
recalibration of the images using the most up-to-date reference files and
the IRAF task {\tt calstis}.  Figures~\ref{47tucimage} and~\ref{m15image}
show the resulting STIS LP images of 47~Tuc and M15, respectively.

\subsection{WFPC2}

WFPC2 consists of four 800~$\times$~800 pixel CCDs -- the Planetary
Camera (PC), which has a plate scale of
0.$^{\prime\prime}$0455~pixel$^{-1}$, and three wide-field (WF) CCDs,
each having an 80$\arcsec$~$\times$~80$\arcsec$ field of view
(0.$^{\prime\prime}$1~pixel$^{-1}$).  A general description of WFPC2
and its on-orbit performance is given by \citet{trauger} and by
\citet{holtz1}.  As Table~\ref{hstdata} indicates, we used WFPC2
observations of 47~Tuc and M15 that were taken with the broad band
F606W and F814W filters.  The former is well approximated as a wide
version of Johnson's V-band filter and the latter produces magnitudes
which are nearly equivalent to Cousins I-band magnitudes, so we will
refer to magnitudes in the F606W and F814W filters as \myv\ and \myi,
respectively.

The WFPC2 observations of M15 have been previously analyzed by
\citet{dmp95a} and by \citet{pck97}.  G. Piotto (private communication)
kindly provided the \myv\ and \myi\ magnitudes and the WFPC2
coordinates of the M15 stars used by \citet{pck97} in their
determination of the luminosity function of this cluster.  These data
have a total integration time of 6050 sec in each of the F606W and
F814W filters.

The WFPC2 data for 47~Tuc were taken from the HST Archive and recalibrated 
using the task {\tt calwp2}, again adopting the most up-to-date
reference files; a subset of these observations, those from HST proposal 5091,
were previously analyzed by \citet{dmp95b}.  The multiple WFPC2 images of
47~Tuc were combined using a basic shift-and-average technique.  The pixel
offsets between images were determined from the positions of 275 relatively
bright, isolated stars and ranged from 0--42 pixels for the thirteen
F606W images; the two F814W images were not offset with respect to one
another.  After aligning the images and scaling by the exposure times,
the images in each filter were averaged, ignoring bad pixels
determined from the CCD characteristics or flagged in the data quality
files.  The combined images, which had total integration times of
5530~sec in F606W and 730~sec in F814W, were then used for the WFPC2
photometry of 47~Tuc.

\section{Photometry}

All of the photometry was performed with tasks within the
{\tt daophot} package of IRAF.  First, {\tt daofind} was used to detect stars
greater than 3.5$\sigma$ above the background on each image.  Stellar 
magnitudes were then calculated with {\tt phot}, using a 2-pixel aperture 
and a sky annulus with an inner radius of 0.$^{\prime\prime}$5 and a width of
0.$^{\prime\prime}$5.
Finally, multiaperture photometry of 40--50 bright, isolated stars was used
with {\tt mkapfile} to determine the aperture corrections needed to transform
the 2-pixel magnitudes to those appropriate for a 0.$^{\prime\prime}$5 aperture;
the resulting aperture corrections are given in Table~\ref{magcorrtable}.

\subsection{STIS Photometry}

For the STIS LP observations, 1307 and 1041 objects were detected by
{\tt daofind} in the 47~Tuc and M15 images, respectively.
The top panels of Figure~\ref{magerrs}
show the 1$\sigma$ uncertainties in the measured (2-pixel aperture)
\mystis\ magnitudes of these ``stars'' as a function of magnitude.
The greater scatter in the 47~Tuc uncertainties
is due to the larger number of image artifacts (diffraction
spikes, etc.) seen in the 47~Tuc image
(Figure~\ref{47tucimage}) and detected as
``stars'' by {\tt daofind}; these are eliminated in later processing and
analysis of the data.

\subsection{WFPC2 Photometry}

As mentioned previously, the WFPC2 magnitudes of the M15 stars were provided
by G. Piotto; the corresponding data reduction and photometric techniques,
which differ somewhat from our methods (see Section 5), are
discussed fully by \citet{pck97}.  We utilized only their
WF4 data because our STIS image of M15 falls entirely within its field of view.

For 47 Tuc, the WFPC2 data from the WF2 chip completely encompasses
our STIS image, and {\tt daofind} detected 4206 and 3886 ``stars'' on
the combined WF2 F606W and F814W images, respectively.  The bottom
panels of Figure~\ref{magerrs} show the 1$\sigma$ uncertainties in the
measured (2-pixel aperture) \myv\ and \myi\ magnitudes of these
stars as a function of magnitude.
 
WFPC2 data have been shown to be subject to three effects that can
produce systematic errors in the photometric magnitudes measured with
this instrument: geometric distortion, charge transfer efficiency
(CTE) effects, and the long-versus-short (LvS) anomaly.  The geometric
distortion of WFPC2 and the CTE effect are fully discussed by \citet{holtz1} and
\citet{holtz2} and in the latest versions of the HST Data Handbook and the
WFPC2 Instrument Handbook \citep{wfpc2book}; the latter two references also
discuss the LvS anomaly.  A brief description of each of these effects follows.

Geometric distortion causes the effective pixel size to be smaller at the edge
of a WFPC2 chip than at its center, making the integrated
magnitude of a star appear progressively fainter as it falls further
from the center of the chip.  For a CCD read out from top to bottom
and left to right, the CTE effect causes a star to have
a fainter measured magnitude when it falls near the top (and/or
left-hand side) of one of the WFPC2 CCDs than when it falls near the
bottom (and/or right-hand side) of the same chip.  This is caused by
charge trapping, i.e. loss of charge, as the signal from the star is
transferred across the chip during readout.  The CTE loss is greatest
for faint stars with low background levels, and the top-to-bottom loss
is generally larger than the left-to-right loss.  For faint stars, the
CTE effect has also become more pronounced with time.  The LvS anomaly
produces stellar magnitudes which are brighter in longer exposures, or
more correctly, which appear brighter as their total count levels
increase; its cause is currently unknown (and its existence has recently been
questioned by Dolphin 2000). 

We have not corrected the WFPC2 magnitudes of 47~Tuc for any of the three
effects described above.  Correcting for
geometric distortion involves
multiplying the WFPC2 image by a correction image before performing photometry
(see the HST Data Handbook).
However, since geometric distortion mainly affects stars lying near
the edges of the WFPC2 chips, while our STIS frames tend to overlie the central
part of the respective WF images and cover only about 22\% of the field-of-view
of a WF camera, we have ignored this correction.  The latest correction
algorithms for the CTE effect and the LvS anomaly are described by
\citet{whitmore} and \citet{cm98}, respectively, but each depends upon the
total counts detected for a star and are thus exposure-time dependent and
inapplicable to data made up of a combination of images having different
exposure times.  For this reason, and because the WFPC2 photometry of M15 was
not corrected for these two effects by \citet{pck97}, we have not corrected the
WFPC2 magnitudes of the 47~Tuc stars for either the CTE effect or the LvS
anomaly.  Nevertheless, we have estimated how these two phenomena would affect
our measured WFPC2 magnitudes and have used these estimates to limit the
magnitude range over which we analyze our photometry; we will discuss these
magnitude selection criteria further in Section 3.5.

\subsection{Flight-System Magnitudes}

We define a flight system magnitude, FMAG, similar to \citet{holtz2}:

\begin{equation}
{\rm FMAG} = -2.5 \times {\rm log}\ ({\rm DN\ s^{-1}}) + {\rm ZP} + 2.5 \times {\rm log\ GR},
\end{equation}

\noindent{where DN are counts within a 0.$^{\prime\prime}$5 aperture, ZP is
the photometric zero point correction, and GR is the gain
ratio used for the given CCD.  \citet{holtz2} defined GR to be unity for WFPC2
observations made with {\tt gain}=14; we define GR to be unity for STIS
CCD observations made with {\tt gain}=1.
Table~\ref{magcorrtable} lists the zero point corrections, gain ratios, and
aperture corrections used here to put our 2-pixel aperture photometry onto the WFPC2 and STIS flight systems.}

\subsection{Matching Stars in F606W and F814W}

The positions of 270 of the stars used to align the F606W images of 47~Tuc
were used with the IRAF task {\tt imagfe} to
determine the coordinate transformation between the combined F606W and
combined F814W images of 47~Tuc.  Using this transformation, the coordinates
of the stars detected in F814W were matched to those detected in F606W,
accepting only those matches for which the predicted and measured positions
agreed within 1.0 pixel.  This produced a list of 2605 stars having both
\myv\ and \myi\ magnitudes in the WF2 frame of 47~Tuc.  The M15 data from
\citet{pck97} contained 3465 WF4 stars detected in both
the F606W and F814W filters.

The top panels of Figures~\ref{47tucwfpc2cmds} and~\ref{m15wfpc2cmds} show
the resulting WFPC2 color-magnitude diagrams (CMDs) of these clusters; the
magnitudes used in these figures have now been corrected to a 0.$^{\prime\prime}$5
aperture.  Figure~\ref{47tucwfpc2cmds} shows a tight main sequence for 47~Tuc,
with the main-sequence turnoff occuring near the bright end of the data
displayed there.  A second sequence of stars is also visible in this figure,
lying about 1 magnitude bluer than the 47~Tuc main sequence for \myv~$>$~22;
this is composed of main-sequence stars in the Small Magellanic Cloud (SMC),
against which 47~Tuc is projected on the sky.  Figure~\ref{m15wfpc2cmds}
shows a similarly well-defined main sequence for M15, with the
main-sequence turnoff again appearing at the brightest magnitudes shown.

\subsection{Selecting the Main-Sequence Stars}

For the comparisons between the WFPC2 and STIS photometry of M15 and 47~Tuc,
we restricted our analysis to stars that: 1) lie on the main sequences of these
clusters, and 2) have photometry that has not been significantly affected
by the CTE effect or by the LvS anomaly discussed in Section 3.2.  Using the
correction formula for the LvS anomaly given by \citet{cm98}, we determined that
the photometric magnitudes would become systematically too faint by about
0.1~mag at \myv~$\sim$~24.5 and \myi~$\sim$~23 for 47~Tuc and at \myv~$\sim$~25
and \myi~$\sim$~24 for M15.  Thus, we applied these magnitude cuts to the WFPC2
data before determining which stars were main-sequence members; the magnitude
bounds adopted for 47~Tuc and M15 are shown as dotted horizontal lines in the
top panels of Figures~\ref{47tucwfpc2cmds} and~\ref{m15wfpc2cmds}, respectively.
The boxes described by the solid lines in the top panels of
Figure~\ref{47tucwfpc2cmds} surround the regions of the CMDs used to determine
the location of the SMC main sequence.
There are 2049 and 2057 stars lying between the \myv\ and \myi\ magnitude
limits adopted for 47~Tuc, respectively; 143 of these stars lie within the SMC
box in the \myv\ vs. \mycolor\ CMD, while only 71 lie within the analogous box
in the \myi\ vs. \mycolor\ CMD.  For M15, 1950 and 2043 stars lie within the
chosen \myv\ and \myi\ bounds, respectively.
We estimate that the CTE
effect will cause the fluxes that we measure for the faintest stars used
in the main-sequence fitting to be about 3\% too faint (on average) while also
introducing an additional scatter of about 0.03 mag in the WFPC2 magnitudes
measured for the faintest stars.

After applying the magnitude cuts shown in Figures~\ref{47tucwfpc2cmds}
and~\ref{m15wfpc2cmds}, we selected the main-sequence stars by fitting
ridge lines to the remaining data using an iterative procedure.  First, the
stars were binned in 0.5~magnitude bins, and the median \mycolor\ color and
mean magnitude were calculated in each bin.  A fourth-order polynomial
(second-order for the SMC) was fit to the median color as a function of
magnitude, producing an equation for the main-sequence ridge line.
Then,
in each magnitude bin, the average color deviation from the ridge line was
calculated,
and a second-order relation was fit to the average color deviation as a
function of magnitude.  Using these two relations, we could then predict both
the ridge-line color of every star in the CMD and the number of average
deviations that the measured color of that star differed from the
ridge-line color.
To procure the final group of main-sequence stars in each cluster, we discarded
all stars lying more than three average deviations from the ridge line and
reiterated this procedure until no remaining stars were discarded.  This
algorithm was applied to both the \myv\ vs. \mycolor\ and \myi\ vs. \mycolor\
CMDs of each cluster, and the CMDs of the stars selected to be
main-sequence members of 47~Tuc and M15 are shown in the bottom panels of
Figures~\ref{47tucwfpc2cmds} and~\ref{m15wfpc2cmds}, respectively; the bottom
panels of Figure~\ref{47tucwfpc2cmds} also shows the stars which lie on
the main sequence of the background SMC.  In the end, 1670, 120, and 1756 stars
remained along the \myv\ vs. \mycolor\ main-sequence ridge lines of
47~Tuc, the SMC and M15, respectively.  Likewise, the
\myi\ vs. \mycolor\ ridge lines were defined by 1753, 61, and 1845
stars for 47~Tuc, the SMC and M15, respectively. 

\subsection{Matching Stars in WFPC2 and STIS}

In exactly the same manner in which the F606W and F814W detections for
47~Tuc were matched, the stars detected with WFPC2 were paired with
those seen in the STIS frames.  Here, the positions of the stars used
to calculate the \mystis\ aperture corrections were used to
determine the coordinate transformations between the STIS images and
the corresponding WFPC2 frames of each cluster, and stars which were
located within 1.0 pixel of the expected position on the WFPC2 frames
were matched to those detected with STIS.  CMDs containing the stars
detected in F606W, F814W and STIS LP are shown in the top panels of
Figures~\ref{47tucallcmds} and~\ref{m15allcmds} for 47~Tuc (and the SMC) and for
M15, respectively.  There were 657 stars detected in all three filters
in 47~Tuc and the SMC; 796 stars were found in all three filters in M15.  The
bottom panels of Figures~\ref{47tucallcmds} and~\ref{m15allcmds} show
the subsets of the 47~Tuc and M15 stars, respectively, which also lie along
the WFPC2 main sequences shown in the bottom panels of
Figures~\ref{47tucwfpc2cmds}
and~\ref{m15wfpc2cmds}; these samples are made up of 410, 19 and 467
stars in 47~Tuc, the SMC and M15, respectively, and are the stars
that we used to determine the transformations between the \myv\ and
\myi\ magnitudes measured with WFPC2 and the \mystis\ magnitudes
measured with STIS.

\section{Transformations Between the STIS and WFPC2 Systems}

Before computing the transformations from \myv\ and \myi\ to \mystis,
the flight magnitudes were corrected for extinction, using the reddening values
for the given line of sight to each globular cluster; these reddenings 
(taken from the on-line globular cluster catalog described by
Harris 1996) and the resulting extinction corrections are listed in
Table~\ref{magcorrtable}.  The reddening-corrected color-color diagrams of
the ridge-line stars detected in both WFPC2 filters and with the
STIS LP filter are shown in Figure~\ref{colorcolor}, where the 47~Tuc and SMC
data are plotted in the left-hand panels as circular points and asterisks,
respectively, and the M15 data are plotted in the right-hand panels.

For each
instrument/filter combination, the appropriate extinction corrections were
derived using SYNPHOT to calculate magnitudes as a function of
reddening for the five spectra of K5~III stars listed in the
Bruzual-Persson-Gunn-Stryker (BPGS) Spectrophotometry Atlas,
adopting the reddening law of \citet{cardelli}.  At a given reddening, the
magnitudes of the five stars were averaged and the difference between this
mean magnitude and that measured at zero
reddening was adopted for the extinction at that reddening.  For the WFPC2
magnitudes, this procedure also involved averaging the extinctions derived
from each of the three WF chips.  The resulting extinction corrections appear
to be quite dependable, as the A$_{814}$ values differ by only $\sim$0.001 mag
from those determined independently by \citet{holtz2} and \citet{hgvg}, and
the A$_{606}$ values differ by less than 0.010 mag from those estimated by
\citet{hgvg}.

We note that, in principle, extinction corrections can be nonlinear, complex
functions of both the line of sight and the spectrophotometric properties of
the source of interest.  However, the extinction levels in the 47~Tuc and M15
fields are sufficiently small that second-order effects are estimated to be much
smaller than the random photometric errors.  This is supported by alternative
calculations of the extinction corrections, using the method described above
but substituting the three spectra of G8~V stars from the BPGS
Spectrophotometry Atlas for the spectra of the K5 giants.  The corresponding
extinction values are systematically higher than those adopted here, but the
differences are $\lesssim$0.01 mag in each of the three filters considered.

In deriving transformations between the WFPC2 and STIS photometric
systems, we have considered two questions.  First, are the transformations
metallicity-dependent?  Second, what is the functional form of the
transformations?  To address the latter question, we have
performed both linear and quadratic least-squares fitting with
iterative 3-$\sigma$ rejection, so the transformation relations take
the general form:

\begin{equation}
{\rm I}_{\rm LP,0} = {\rm m}_0 + {\rm ZP} + {\rm a} \times ({\rm V}_{606}-{\rm I}_{814})_{0} + {\rm b} \times ({\rm V}_{606}-{\rm I}_{814})^2_{0}
\end{equation}

\noindent{where m$_0$ is the extinction-corrected WFPC2 magnitude, either
V$_{606,0}$ or I$_{814,0}$, and ZP is the zero point of the fit.}

If metallicity affects the color transformations, then we would expect the
color-color relations of 47~Tuc and M15 to differ.  This is indeed the case when
linear relations are fit to the entire samples of the 47~Tuc and M15 ridge-line
stars; these fits are shown as dashed lines in Figure~\ref{colorcolor}, and
the open points in this figure are the data thrown out by the iterative,
3-$\sigma$ rejection during the fitting.  The coefficients of the dashed
relations are listed in the uppermost section of
Table~\ref{combtransforms}, where it can be seen that both the zero points and
the intercepts of the fits to these particular M15 and 47~Tuc photometric data
differ by more than 3$\sigma$.

Does this imply that there is a metallicity dependence in the color
transformations?  Not necessarily.  Due to the different metallicities and
distance moduli of these two clusters, the range of V$_{606}$--I$_{814}$ colors 
which our selection criteria allow are not the same in each.  As
Figure~\ref{colorcolor} shows, the 47~Tuc ridge-line stars
have 0.6 $\lesssim$ V$_{606}$--I$_{814}$ $\lesssim$ 1.8, while those in M15
have 0.3 $\lesssim$ V$_{606}$--I$_{814}$ $\lesssim$ 1.0.  The M15 data extends
to bluer colors because M15, with [Fe/H]=--2.25 \citep{harris}, is more
metal-poor than 47~Tuc, which has [Fe/H]=--0.76 \citep{harris}.  However, since the apparent
distance modulus of M15 is $\sim$2 magnitudes greater than that of 47~Tuc, the limiting
magnitude of the STIS observations, coupled with the magnitude limits that we
apply to the WFPC2 data, allows the 47~Tuc stars to extend to much redder
V$_{606}$--I$_{814}$ colors than those in M15.  The significance of these color
range differences is that any curvature
or break in slope of the true transformation relations could affect the linear
fits to the two clusters differently.  Thus, we must compare stars of similar
color in 47~Tuc and M15 to look for metallicity effects. 

Considering only the stars with 0.6 $\leq$ V$_{606}$--I$_{814}$ $\leq$ 1.0 in
each cluster (those lying between the dotted, vertical lines in
Figure~\ref{colorcolor}), linear least-squares fits produce the relations shown
as solid lines in Figure~\ref{colorcolor}.  The middle section of
Table~\ref{combtransforms} shows that these new transformation relations
for 47~Tuc and M15 now differ at only the 1$\sigma$ level.  This improved
agreement is
caused by the substantially different fit for the color-restricted set of
47~Tuc stars than for the entire 47~Tuc sample; the M15 transformations do not
change appreciably when the color boundaries are applied.  Thus, metallicity
effects are apparently unimportant in the transformation relations, 
and either
the true transformations contain higher-order terms, or there is a change in
slope at some color redder then V$_{606}$--I$_{814}$ = 1.0.

Since there is no obvious sudden change in slope of the 47~Tuc color-color
data at any color, higher-order fits are suggested.
We have chosen to represent the WFPC2-to-STIS transformation equations by
quadratic, least-squares fits.  The fact that there is no metallicity
dependence also allows us to  combine all of the 47~Tuc, SMC and M15
data.  The final transformations are illustrated in Figure~\ref{finaltrans},
where the 47~Tuc, SMC
and M15 stars are shown as circles, triangles and squares, respectively, and
the quadratic fits are the solid curves drawn there.  For clarity, we have not
plotted the stars rejected during the fitting.
The coefficients of these transformation relations are listed in the bottom
section of Table~\ref{combtransforms}.

Figure~\ref{magdevs} shows the differences between the \mystis\
magnitudes measured for the 47~Tuc/SMC stars (left-hand panels) and
the M15 stars (right-hand panels) and the magnitudes predicted by the quadratic
transformation relation shown in the bottom panel of
Figure~\ref{finaltrans} as functions of both \myv\ magnitude (top
panels) and color (bottom panels).  Here, the globular cluster stars
are shown as filled points, and the SMC stars are asterisks; we have again
omitted the stars rejected during the least-squares fitting.  From
this figure, it is apparent that the transformation relations
generally predict \mystis\ magnitudes for the 47~Tuc and M15 stars
within 0.05 mag (1$\sigma$) of those measured, with no systematic offset.
However, the predicted \mystis\ magnitudes of the SMC stars, while also being
good to about 0.05 mag, may be systematically $\sim$0.05 mag too bright.

\section{Discussion}

We have argued that  the differences between the linear fits to the
complete samples of 47~Tuc and M15 color-color data shown in
Figure~\ref{colorcolor} are due to the presence of a higher-order
term in the ``true,'' metallicity-independent transformation relations.
Here, we discuss some other
effects which could also produce this outcome and explain why we feel that
these are  less important.

One possible source of differential systematic errors in the photometry is the
different methodologies used to measure the WFPC2 magnitudes of 47~Tuc and M15.
While both we and \citet{pck97} performed aperture photometry, \citet{pck97}
used point-spread functions (PSFs) to first subtract nearby neighbors, which
should improve the
measured magnitudes, especially for the faintest stars.  In addition,
PSF-fitting allows the removal of spurious
stellar detections (e.g. background galaxies, bad pixels) through various
goodness-of-fit selection criteria, such as $\chi^2$ and sharpness.
However, since the 47~Tuc and M15 fields observed here are not particularly
crowded, the photometric improvements produced by PSF-fitting are only expected
to be significant at the faintest magnitudes.  Our use of magnitude cutoffs
for the WFPC2 data and our exclusion of objects which do not lie along
the cluster main-sequence ridge lines have likely minimized any differences
caused by the photometric algorithms employed.

Likewise, the magnitude cutoffs that we have adopted should preclude any
significant biases in the photometry due to incompleteness of the WFPC2 data.
For the WF3 observations of M15,
\citet{pck97} estimated their data to be 93\% complete at \myv=25 and
92\% complete at \myi=24; these levels should be approximately the same for the
WF4 data, for which \citet{dmp95a} give a completeness level of 92\% at \myi=24
and \myi--\myv$\sim$1.
We also expect the WF2 image of 47~Tuc to be more than 90\% complete at
the magnitude cutoffs used here.  Although \citet{dmp95b} estimated their WF2
completeness level to be only $\sim$80\% at \myi=23 and \myi--\myv$\sim$1.65, this
incompleteness was dictated by their F606W data.  We use a much more extensive
set of F606W observations than they did, so our completeness level will be
determined by the F814W data, which \citet{dmp95b} show to be $\sim$80\%
complete one full magnitude fainter than our \myi\ cutoff.

Could the documented  complications with WFPC2 photometry be the cause of the different
linear transformation relations derived from the complete samples of 47~Tuc and
M15 ridge-line stars?
As discussed previously, the WFPC2 magnitudes in both clusters will be
systematically too faint (about 3\% on average) due to the CTE effect, but the
differential
effect between the two clusters should be negligible.  The LvS
anomaly, on the other hand, while also causing the WFPC2 magnitudes to be
systematically too faint, will become increasingly important as the magnitudes
become fainter.  To first order, the errors in \myv\ and \myi\ will offset one
another, so the \myv--\myi\ colors will not be as greatly affected as the
\myv--\mystis\ and \mystis--\myi\ colors will be.  While we have attempted to
compensate for the LvS anomaly through the magnitude limits that we
have applied to the WFPC2 data, the 47~Tuc and M15 stars will still have
\myv--\mystis\ colors which are progressively too red and \mystis--\myi\ colors
which are progressively too blue as fainter stars are considered.
This will introduce a slight bit of curvature in the color-color relations
shown in Figure~\ref{colorcolor} and could potentially account for
the differing transformation relations derived for the two clusters.  However,
this ``curvature'' will not occur at the same color in 47~Tuc and M15 (because
of their differing distances and metallicities) and therefore should affect the
fits derived from the color-restricted samples of their members as well.  Since
this is not seen, we do not expect that the CTE effect and LvS anomaly make 
significant contributions to the resulting transformation equations.

We have quantified  this further by performing photometry on the longest-exposure
F606W and I814W images of 47~Tuc listed in Table~\ref{hstdata}.  We then
used the algorithms suggested by \citet{whitmore} and by \citet{cm98}
to correct the measured counts for the CTE effect and for the
LvS anomaly, respectively,
and calculated new magnitudes for these stars.  Without applying any magnitude
cutoff to these data, the same analysis procedure used for the combined
images was adopted, resulting in a new determination of
the 47~Tuc ridge-line stars.  This set of data contains a smaller
number of stars than the data resulting from the combined WFPC2 images
because, even though it extends to slightly redder colors than the
magnitude-limited, combined data, most of the bright, blue main-sequence
stars are no longer included because they are saturated in the 1500-sec F606W
exposure.

Performing least-squares fits (with iterative 3-$\sigma$ rejection) to the
corrected, single-image colors of the 47~Tuc ridge-line stars, the resulting
linear and quadratic transformation relations have the coefficients listed in
Table~\ref{singtransforms}.  Comparing these coefficients to the analogous fits
from Table~\ref{combtransforms}, we see that the linear fits have nearly
identical slopes, while their intercepts differ by about 3\%, as expected from
the CTE effect.  Thus, the magnitude cutoffs that we have applied to the
combined-image data appear to have reasonably precluded the LvS anomaly from
significantly biasing our results.  This is reinforced by the 1$\sigma$
agreement between the coefficients of the quadratic relations derived from the
combined 47~Tuc, M15 and SMC data (Table~\ref{combtransforms}) and those
derived from the single-image 47~Tuc data alone (Table~\ref{singtransforms}).

\subsection{A Cautionary Note}

Note that the transformation relations that we present
are only valid for magnitudes in a 0.$^{\prime\prime}$5 aperture.
In particular, our transformations differ significantly from synthetic
calculations, and we discuss this further in Appendix A, where we also describe
a possible dependence of the \mystis\ aperture corrections on color.  We
especially encourage those who use SYNPHOT to calculate color-color
transformations to read Appendix A and to use care when applying the
SYNPHOT (infinite-aperture) results to infer observed magnitudes.

\section{Summary}

An important application of multiband photometry is the derivation of
transformation relations between the corresponding photometric
systems.  In this paper, we have used our STIS observations of two
Galactic globular clusters, 47~Tuc and M15, in combination with archival WFPC2
data of these clusters, to derive transformations which can be used to
calculate STIS \mystis\ magnitudes from WFPC2 \myv\ and \myi\ data
for main-sequence stars with colors in the range $0.3 \lesssim$
\mycolor\ $\lesssim 1.8$ and metallicities in the range
$-2.2 \lesssim {\rm [Fe/H]} \lesssim -0.5$.

Considerable effort was spent to choose an optimal sample of stars to use
in deriving the photometric transformations.  This involved applying magnitude
limits to the WFPC2 data to minimize the influences of the CTE effect and the
long-vs.-short anomaly, each of which systematically affect WFPC2 photometry.
We also used least-squares fitting to determine main-sequence ridge lines in
the WFPC2 color-magnitude diagrams of each cluster and rejected those stars
lying more than three average deviations from these ridge lines.

Comparing linear, least-squares fits to the color-color data of 47~Tuc
and M15 over a similar range in \mycolor, we found that the coefficients of
the fits differed by less than 1$\sigma$, indicating that the ``true''
color-color relations have no metallicity dependence.   We therefore combined
the photometry of the 47~Tuc and M15 main-sequence stars (as well as a few
stars in the SMC) and performed quadratic, least-squares fits to this
combined data set to derive the suggested transformation relations.  These
can be used to predict \mystis\ magnitudes from \myv\ and \myi\ magnitudes,
and the agreement between the predicted and measured \mystis\ magnitudes is
generally good to 0.05 mag (1$\sigma$).

\acknowledgments

We thank G.~Piotto for providing the WFPC2 photometry of M15 and for helpful
suggestions regarding its use.   We also express our gratitude to S.~Feltzing,
M.~Bessell, H.~Ferguson, J.~Gallagher, B.~Mobasher,
N.~Tanvir and T.~Smecker Hane for their scientific insights and discussions.
Support for this work was provided by NASA grant
number GO-7419 from STScI, operated by AURA Inc, under NASA
contract NAS5-26555.

\clearpage

\appendix

\section{Synthetic Transformation Relations}

M. Bessell (private communication) has pointed out to us that his synthetic
\myv, \myi, and \mystis\ photometry predicts
different transformation relations than those that we present here.  While his
calculations of the transformations shown in Figure~\ref{finaltrans} verify
that these relations are metallicity-independent and become nonlinear when
stars redder than \mycolor~$\sim$~1.2 are included,
he finds that the coefficients of the synthetic relations differ significantly
from those given in Table~\ref{combtransforms}.  
In this appendix, we compare the results of SYNPHOT synthetic color
calculations with our empirical data to evaluate two possible explanations
for the observed differences: 1) use of an inappropriate
STIS LP filter response function in SYNPHOT, and 2) a color dependence of the
empirical \mystis\ aperture corrections.

\subsection{SYNPHOT Calculations and the STIS LP Filter Response}

We have used SYNPHOT to repeat Bessell's investigation and have confirmed his
basic finding that our empirical transformation relations differ from synthetic
predictions.  To do this, we
calculated \myv, \myi, and \mystis\ magnitudes for stars no.~14--71 (B9--M8
dwarfs) of the BPGS Spectrophotometric Atlas within SYNPHOT.  Ideally, we would
expect only a constant offset between the observed and synthetic magnitudes
because different aperture sizes apply for each --
the SYNPHOT magnitudes are ``infinite-aperture'' magnitudes, while our reported
magnitudes represent the light falling within a 0.$^{\prime\prime}$5 aperture.
However, the \mycolor\ data should not be affected by the differing apertures;
\citet{holtz2} estimate that an aperture correction of --0.1 mag is appropriate
to adjust a 0.$^{\prime\prime}$5 aperture magnitude to an
infinite-aperture magnitude for any WFPC2 magnitude.  The analogous aperture
correction for \mystis\ magnitudes
is unknown, so we have simply normalized the synthetic and empirical relations
at \mycolor~=~0.3, the approximate blue limit of our observational data, to
more easily compare the slopes of the two relations differentially.
The SYNPHOT calculations (open points) are compared to our final (normalized)
transformation relations
in the left-hand panels of Figure~\ref{comp2syn}, where we show our empirical
relations as solid lines spanning only the range of colors used in their
derivation.
The empirical and synthetic color-color relations differ significantly in slope,
diverging by $\sim$0.2 mag between \mycolor~=~0.3 and \mycolor~=~1.8.

Bessell has pointed out that the empirical transformations can be reproduced
by synthetic color-color relations calculated with a modified STIS LP
response function that has a significantly bluer effective wavelength than
that reported in the STIS Instrument Handbook \citep{stisbook}.
To confirm this, we have performed
SYNPHOT color calculations analogous to those described above but adopting
the modified STIS LP filter response suggested by Bessell.  The results are compared to
our final transformation relations in the right-hand panels of
Figure~\ref{comp2syn}.  In these panels of the figure,
the empirical relations are the same as those which appear in the corresponding
left-hand panels, but we have shifted the \mystis\ magnitudes calculated with
the modified filter response
to normalize them with our relations at \mycolor~=~0.3, since the \mystis\ zero
point for the modified filter response will differ from that which we have
adopted (Table~\ref{magcorrtable}).  
The normalized, synthetic colors computed with the modified LP filter
response are quite similar to those observed over the entire range of colors
used in deriving our transformation relations.

To give the reader a feel for the degree to which the STIS LP filter response
would have to be modified to reproduce our empirical transformations,
Figure~\ref{filters} compares the STIS LP filter response tabulated in the STIS
Instrument Handbook for Cycle 10 (Leitherer et al. 2000; dotted line) to that
suggested by Bessell (2000, private communication; solid line).
The transmission profile of the STIS LP filter is thought to be
well-determined (H. Ferguson 2000, private communication), so any possible
discrepancy between the tabulated response function and the ``true'' one would
have to be due to a fairly dramatic change in some other component of the
optical system (e.g., degradation of the CCD sensitivity at longer wavelengths)
to change the filter response to the extent shown in Figure~\ref{filters}.

\subsection{Aperture Corrections}

The PSF of the STIS CCD is known to vary with wavelength, exhibiting an
increasingly pronounced halo at wavelengths longer than about 7500 \AA\ due
to the scattering of light within the CCD \citep{stisbook}; the red light in the
PSF can be scattered to as much as 5$^{\prime\prime}$ from the PSF center.
Therefore, the slopes of our transformation relations may
differ from the SYNPHOT predictions because the fraction of light scattered to
radii greater than 0.$^{\prime\prime}$1 differs for the red and blue
cluster stars that we have observed.  If this is the case, we would need to
apply larger aperture corrections to the magnitudes of the red stars than
those used for the bluer stars to reproduce the SYNPHOT calculations.

We have looked at the behavior of the \mystis\ aperture corrections for stars
of different \mycolor\ color by calculating magnitudes as a function of aperture
radius for: 1) the stars in 47~Tuc that belong to the sample used to derive
our transformation relations, 2) three bright PSF calibration stars for
which HST archive STIS LP observations were available, and 3) Tiny Tim
PSFs generated for dwarf stars of differing spectral types.
Note that the Tiny Tim PSFs do not include the extended halo due to scattered
light, so they will give only a baseline prediction for aperture corrections as
a function of color.

The calibration stars (and HST
proposals from which the archive STIS LP data were taken) are
GRW~+70~5824, a white dwarf (HST proposal 8422), SAO~255271, an F8 star (HST
proposal 8842), and BD~--11~3759, an M4 star (HST proposal 8422).  Since no
WFPC2 observations exist for these stars, their \mycolor\ colors were derived
from the following sources, using transformations from \citet{holtz2} as needed
to convert from V--I to \mycolor: 1) \mycolor~$\sim$~--0.1 for GRW~+70~5824
from Figure 6 of \citet{holtz2}, 2) \mycolor~$\sim$~0.5 for SAO~255271 from an
average of the SYNPHOT colors of F8 dwarfs in the BPGS Spectrophotmetric Atlas,
and 3) \mycolor~$\sim$~2.0 for BD~--11~3759 from an observed (V--I)$_{\rm C}$
color (Leggett 1992).
The Tiny Tim PSFs were derived from the following stars (and their
spectral types) in the BPGS Atlas: star 14 (A0 V), star 41 (F8 V),
star 53 (G8 V), star 65 (K7 V), and star 68 (M3 V); the \mycolor\ colors of
these stars were calculated with SYNPHOT.

We first searched for a color dependence in the aperture corrections that
we have applied to our 2-pixel magnitudes of the 47~Tuc stars to convert them to
magnitudes in a 0.$^{\prime\prime}$5 aperture.  To increase the number of
cluster stars
for which 0.$^{\prime\prime}$5 magnitudes could be measured, we calculated
an empirical PSF based upon 55 stars in the 47~Tuc image.  We then used this PSF
to subtract all of the stars from the image except the star of interest before
measuring its magnitude.  The resulting aperture corrections, shown as open
circles in Figure~\ref{apcorrs}, exhibit substantial scatter for the reddest
stars but do not show an obvious trend with color.  However, the HST
calibration stars hint that the aperture corrections may be larger for redder
stars; these stars are shown as filled points in Figure~\ref{apcorrs}.
Measuring magnitudes in exactly the same manner described in Section~3.1, the 
two blue HST calibration stars (GRW~+70~5824 and SAO~255271) have aperture
corrections of --0.466 and --0.444 mag, respectively, which are in accord
with our estimate (Table~\ref{magcorrtable}).  The aperture correction for
BD~--11~3759, however, is about 0.1 mag greater than this.  This indicates
that the color sensitivity of the aperture corrections listed in
Table~\ref{magcorrtable} is weak at most and is insufficient
to explain the empirical/SYNPHOT differences.  The Tiny Tim
PSFs infer that about 0.05 mag of the different aperture corrections for the
red and blue HST calibration stars is due to effects other than the halo
produced by scattered red light.

Of course, to compare observed magnitudes to SYNPHOT magnitudes, we should
measure magnitudes in an aperture large enough to contain all of the
light of a star.  This is impossible to do in practice, so we have approximated
infinite-aperture magnitudes by magnitudes measured in a 5$^{\prime\prime}$
aperture.  Unfortunately, we could not measure such magnitudes for any of the
47~Tuc stars, even after the other stars were subtracted from the image, because
there was not a single star in the 47~Tuc image which
did not contain at least one bad pixel within its 5$^{\prime\prime}$ aperture.
However, we were able to measure such magnitudes for two
of the HST calibration stars, SAO~255271 and BD~--11~3759, which differ in
\mycolor\ color by about 1.5 mag, and for the Tiny
Tim PSFs, using an aperture with a radius of 5$^{\prime\prime}$ and a sky
annulus having an inner radius of 5$^{\prime\prime}$ and a
width of 0.$^{\prime\prime}$5.  For the two calibration stars, the corrections
needed to transform magnitudes measured in a 0.$^{\prime\prime}$1 aperture to
those in a 5$^{\prime\prime}$ aperture differ by about 0.17 mag, which is
approximately the difference between the empirical and synthetic transformation
relations between \mycolor~=~0.3 and \mycolor~=~1.8, the red end of the
observational data used to
derive the empirical relations.  The Tiny Tim PSFs again indicate that all but
0.05 mag of this difference is due to the scattered-light halo of the PSF.

In conclusion, we cannot exclude the possibility that the differences between
the slopes of the empirical and SYNPHOT transformation relations is
due to a color dependence in the aperture corrections needed to transform the
observed magnitudes to infinite-aperture magnitudes such as those computed
with SYNPHOT.  Thus, the extent that the STIS LP filter response may differ
from that tabulated in the STIS Instrument Handbook for Cycle 10
\citep{stisbook} cannot be determined from the datasets that we have analyzed. 
A more detailed study of the globular cluster fields using PSF-fitting could
potentially prove illuminating but would be complicated by the dependence of
the PSF shape on color and on position on the STIS CCD.
Further investigation of the reasons for the differences between the empirical
and synthetic transformation relations is required (and highly encouraged)
but is beyond the scope of the present paper.

\begin{table}
\dummytable\label{hstdata}
\end{table}

\begin{table}
\dummytable\label{magcorrtable}
\end{table}

\begin{table}
\dummytable\label{combtransforms}
\end{table}

\begin{table}
\dummytable\label{singtransforms}
\end{table}

\begin{figure}[p]
\vspace*{-1.3in}
\hspace*{-1.0in}
\psfig{figure=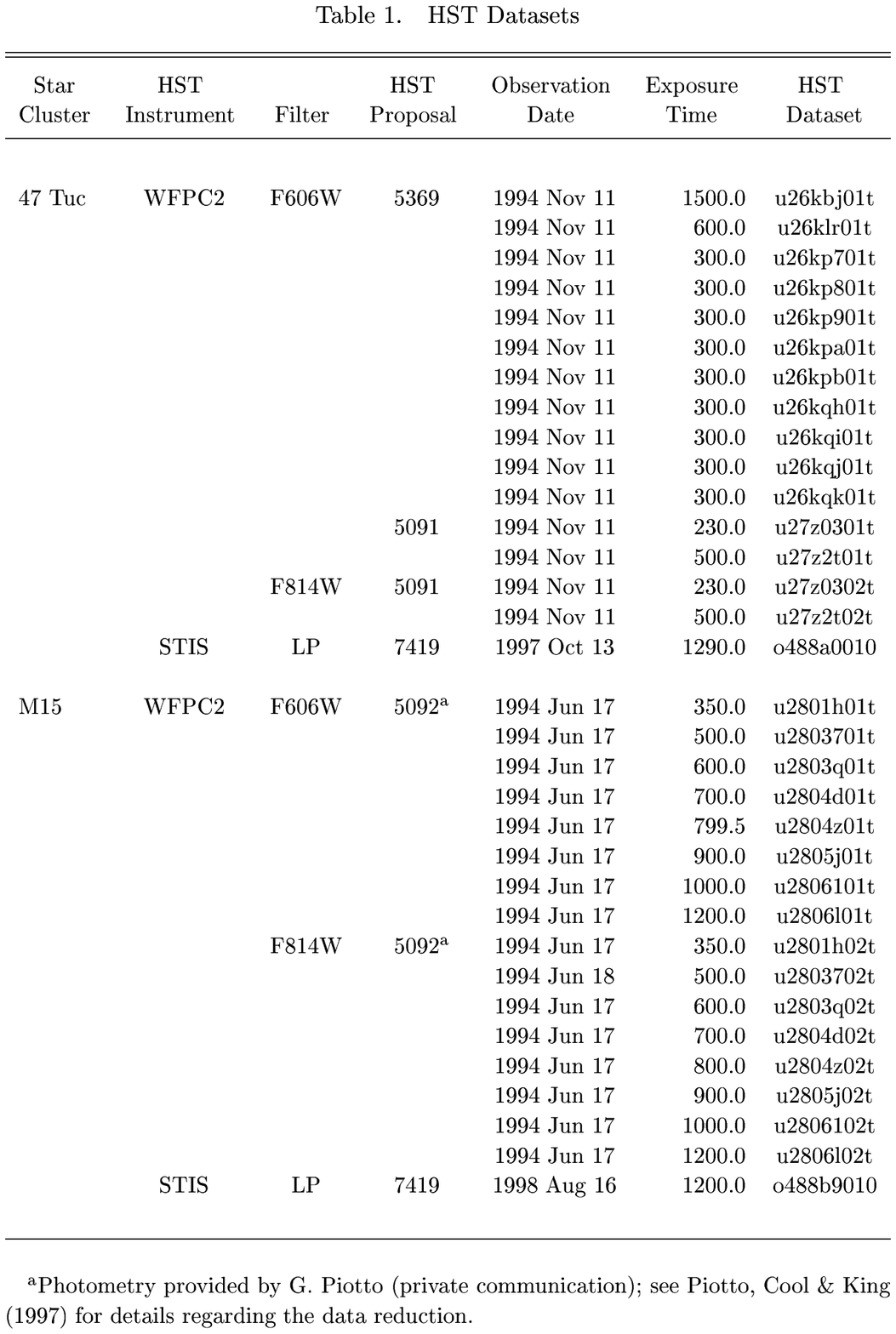}
\end{figure}

\begin{figure}[p]
\vspace*{-1.3in}
\hspace*{-1.0in}
\psfig{figure=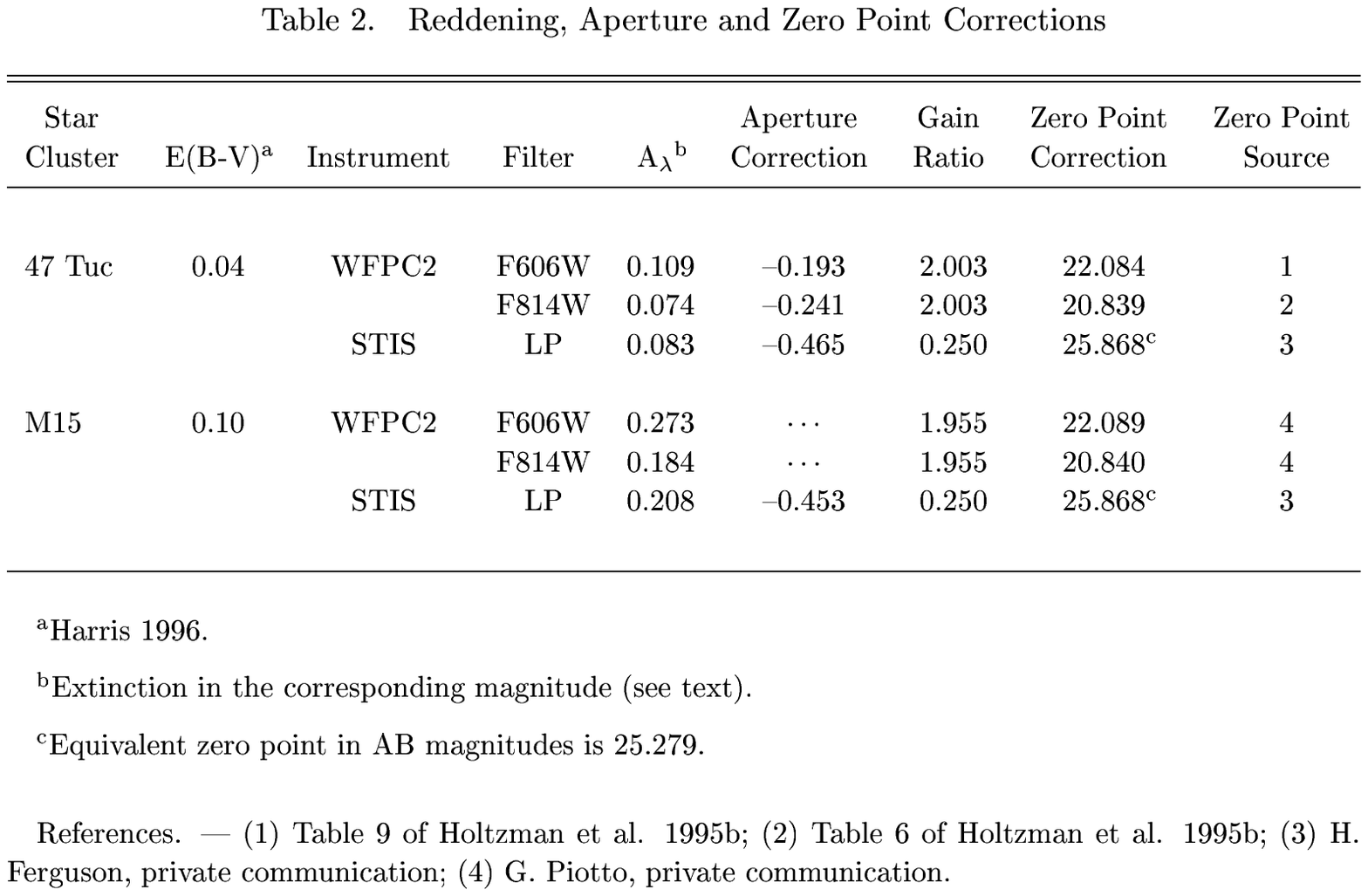}
\end{figure}

\begin{figure}[p]
\vspace*{-1.3in}
\hspace*{-1.0in}
\psfig{figure=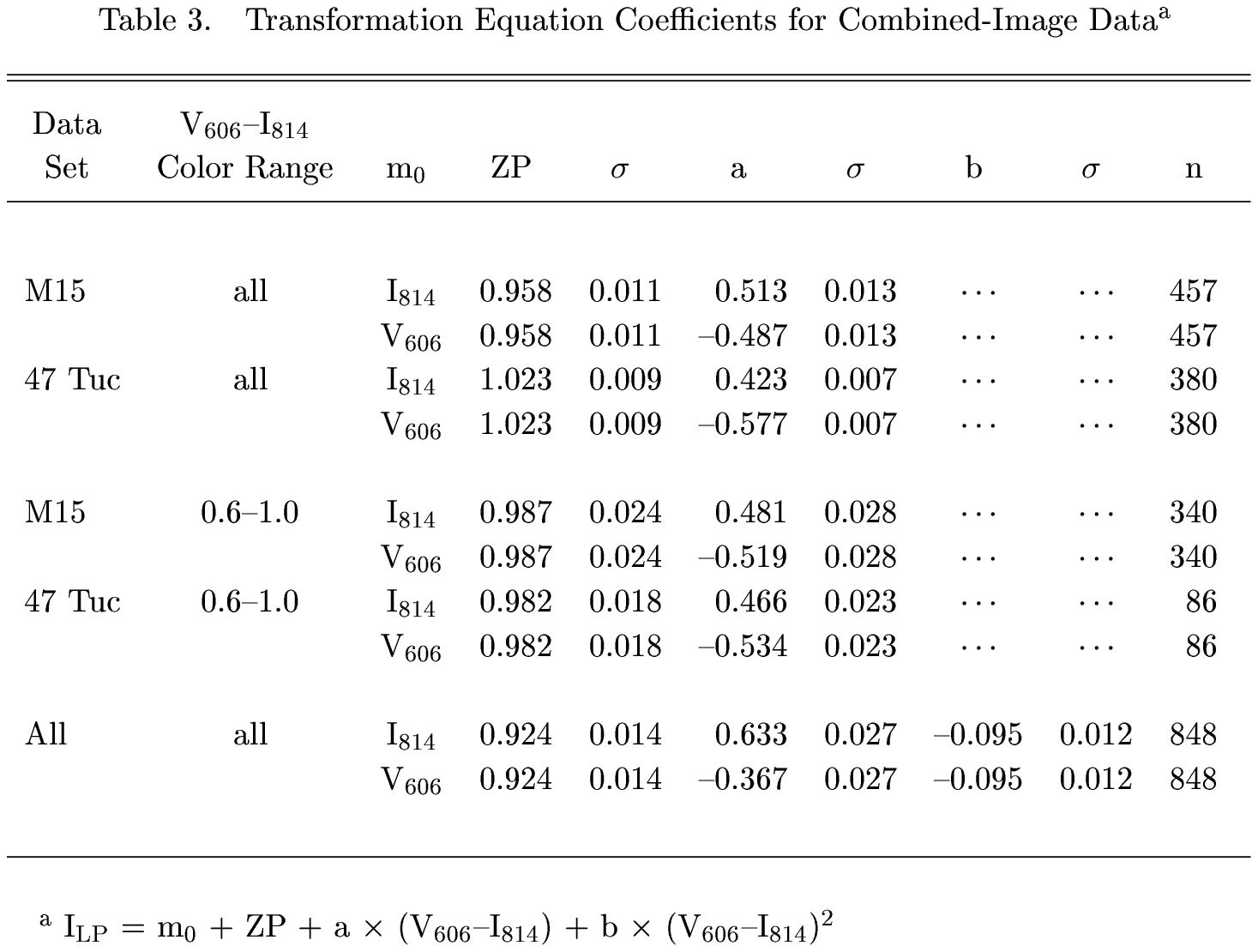}
\end{figure}

\begin{figure}[p]
\vspace*{-1.3in}
\hspace*{-1.0in}
\psfig{figure=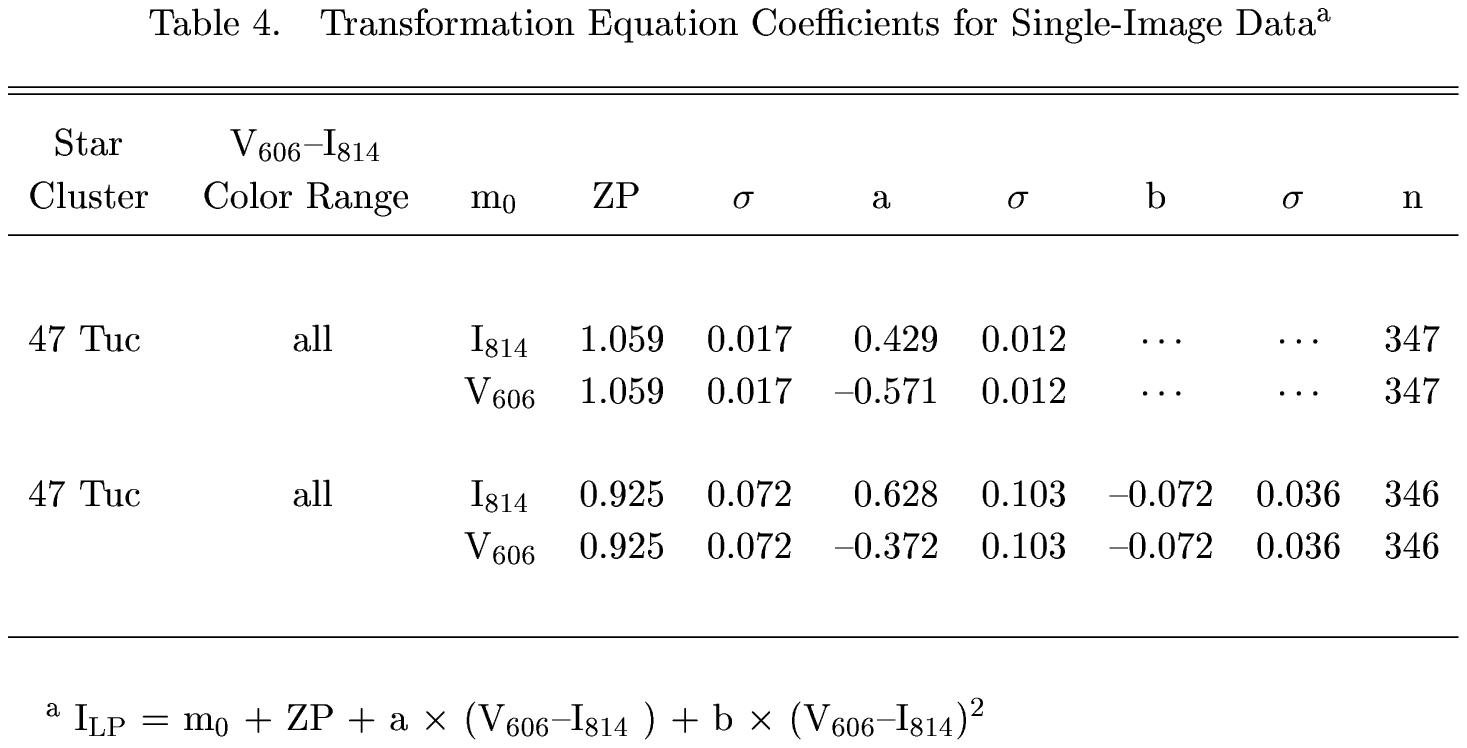}
\end{figure}

\begin{figure}[p]
\vspace*{-0.5in}
\hspace*{0.25in}
\psfig{figure=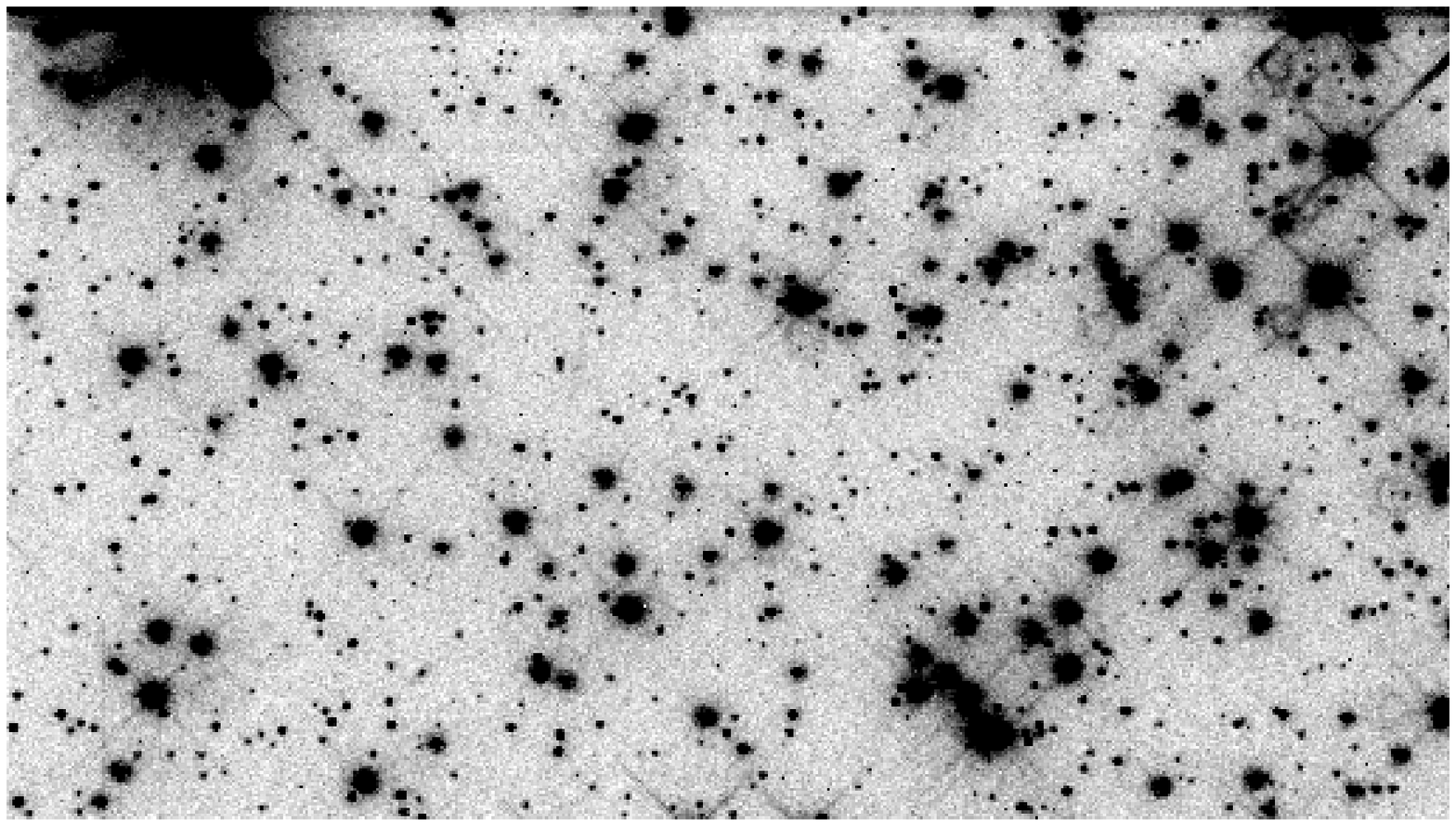,height=6in,width=6in}
\vspace*{-2in}
\caption{STIS LP image of 47 Tuc.  The 28$\arcsec$~$\times$~51$\arcsec$
field of view of the pipeline-processed STIS image of 47~Tuc, taken with the
optical longpass (LP) filter, is shown; this is the image
used to compute \mystis\ magnitudes of the 47~Tuc stars.}
\vspace*{1.9in}
\label{47tucimage}
\end{figure}

\begin{figure}[p]
\vspace*{-0.5in}
\hspace*{0.25in}
\psfig{figure=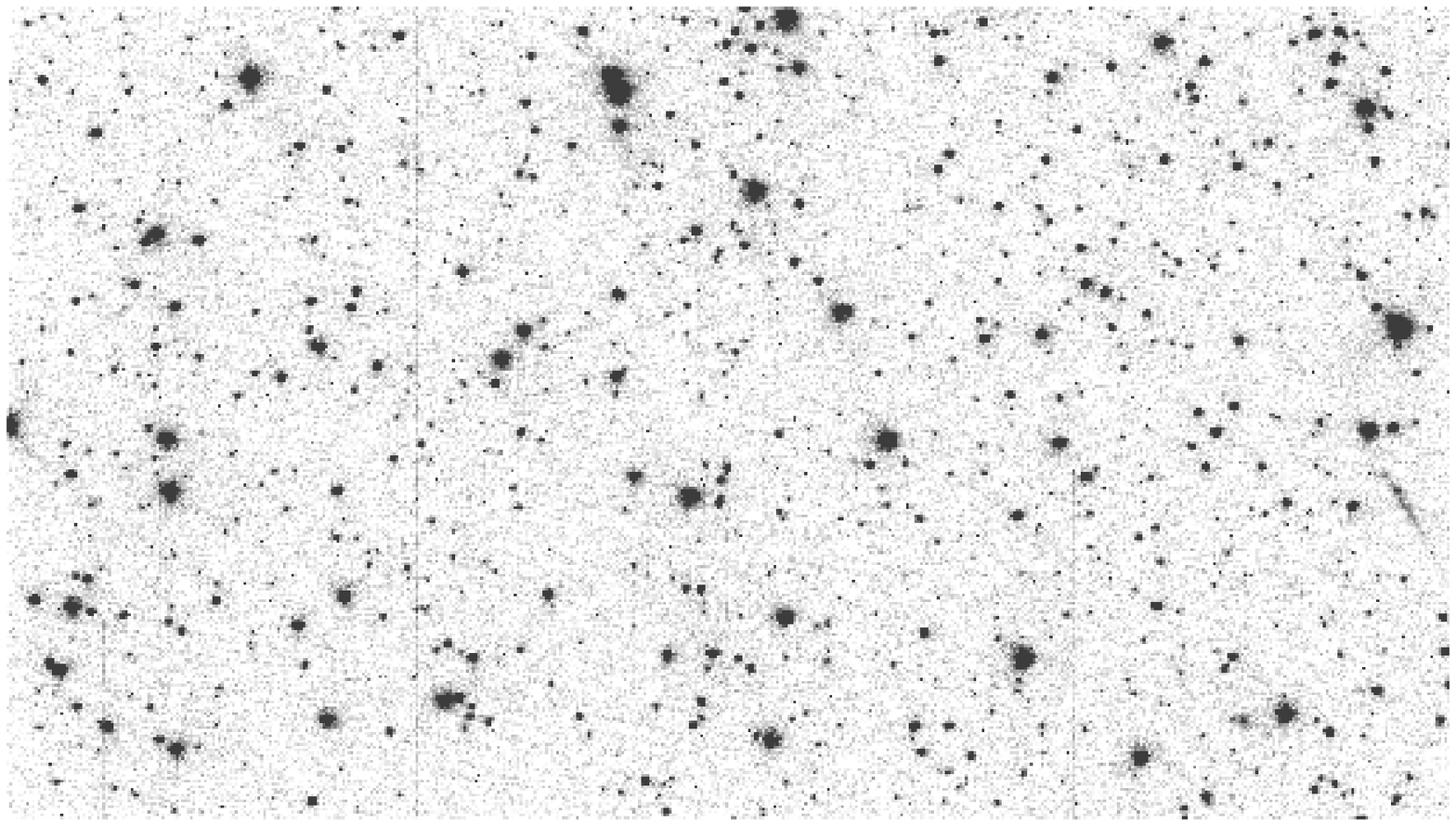,height=6in,width=6in}
\vspace*{-2in}
\caption{STIS LP image of M15.  The 28$\arcsec$~$\times$~51$\arcsec$
field of view of the pipeline-processed STIS image of M15, taken with the
optical longpass (LP) filter, is shown; this is the image
used to compute \mystis\ magnitudes of the M15 stars.}
\label{m15image}
\end{figure}

\begin{figure}[p]
\hspace*{0.25in}
\psfig{figure=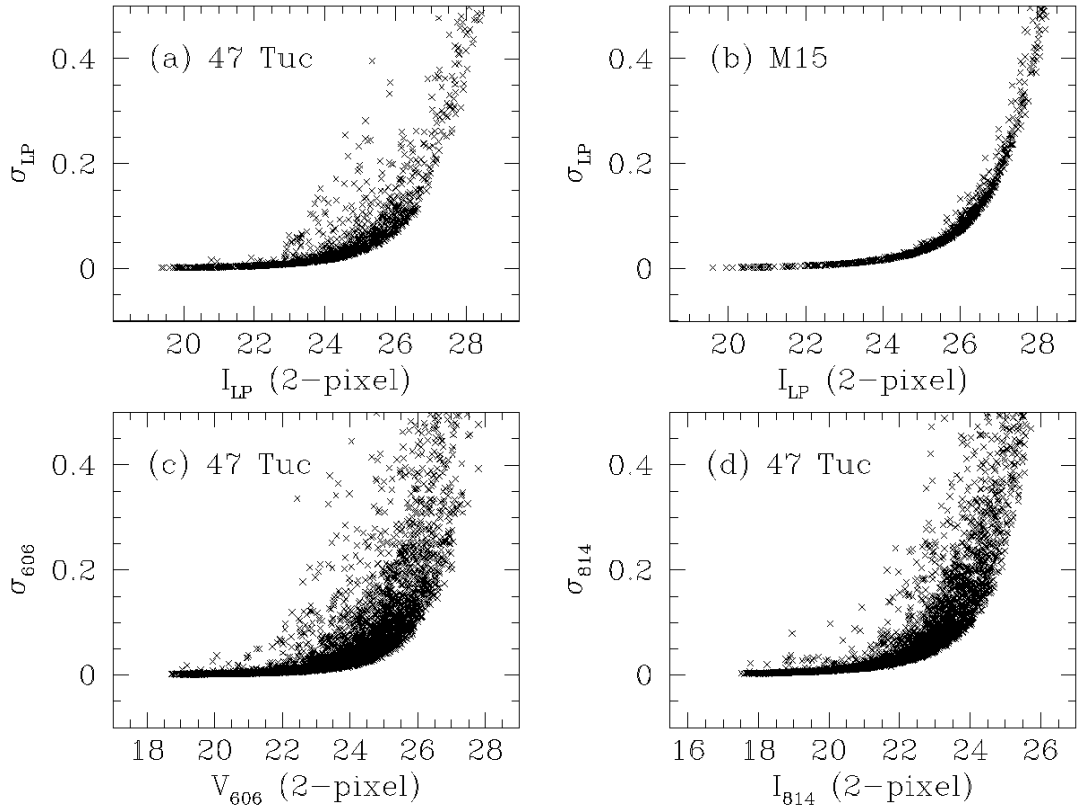,angle=270,height=4.5in,width=6in}
\caption{Photometric uncertainties.  The 1$\sigma$ uncertainties in the
2-pixel aperture photometry are shown as a function of magnitude for (a) the
STIS LP observations of 47~Tuc, (b) the STIS LP observations of M15, (c) the
WFPC2 F606W observations of 47~Tuc, and (d) the WFPC2 F814W observations of
47~Tuc.  The magnitudes plotted do not include the aperture or reddening
corrections given in Table 2.}
\label{magerrs}
\end{figure}

\begin{figure}[p]
\hspace*{0.75in}
\psfig{figure=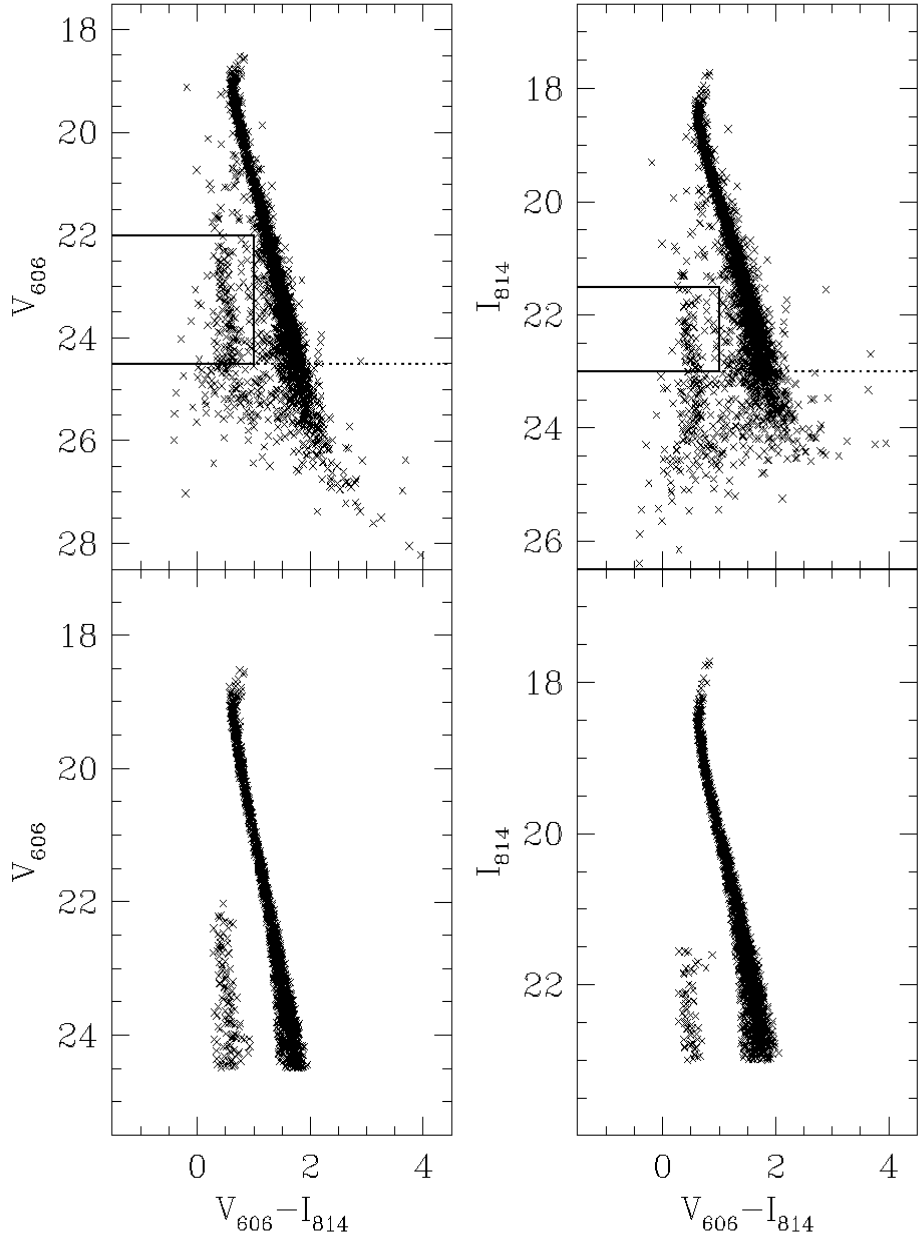,height=7in,width=5in}
\caption{WFPC2 color-magnitude diagrams (CMDs) of 47 Tuc and the Small
Magellanic Cloud (SMC).  The top panels show the CMDs for the 2605 stars
detected
in both the F606W and F814W filters on the WFPC2 WF2 chip.  Dotted lines in
the top panels show the magnitude cutoffs applied before determining
the cluster's main-sequence ridge lines; solid lines show the analogous color
and magnitude limits applied before determining the main-sequence ridge lines
for the SMC.  The bottom panels show those stars determined to lie on the
47~Tuc and
SMC main-sequence ridge lines, which were derived as described in the text.
Reddening and extinction corrections have not yet been applied to the data
plotted here.}
\label{47tucwfpc2cmds}
\end{figure}

\begin{figure}[p]
\hspace*{0.75in}
\psfig{figure=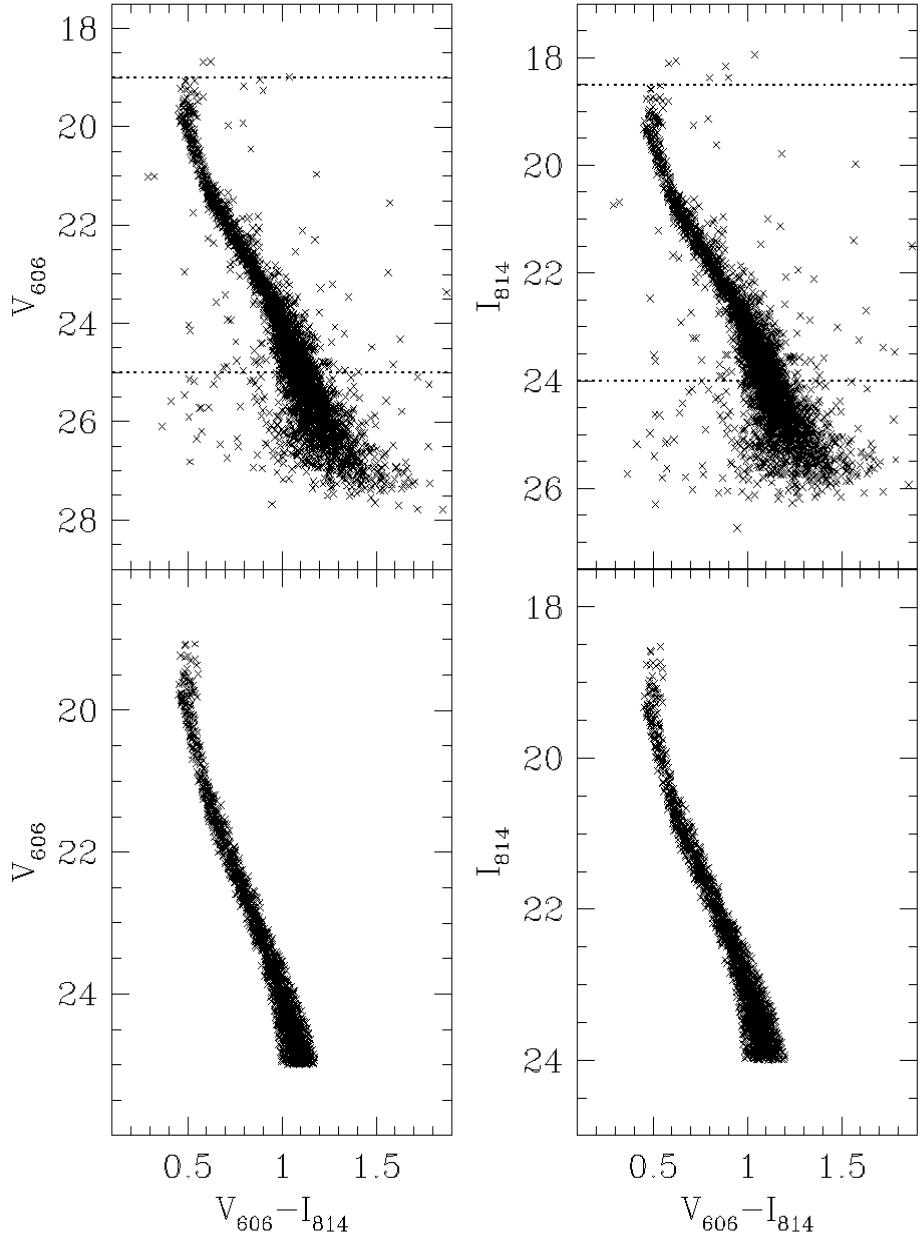,height=7in,width=5in}
\caption{WFPC2 color-magnitude diagrams (CMDs) of M15.  The top panels
show the CMDs for the 3465 stars detected by Piotto, Cool \& King (1997) in
both the F606W and F814W filters
on the WFPC2 WF4 chip.  Dotted lines in the top panels show the magnitude
cutoffs applied before determining the cluster's main-sequence ridge lines.
The bottom panels show those stars determined to lie on the M15
main-sequence ridge lines, which were derived as described in the text.
Reddening and extinction corrections have not yet been applied to the data
plotted here.}
\label{m15wfpc2cmds}
\end{figure}

\begin{figure}[p]
\hspace*{0.25in}
\psfig{figure=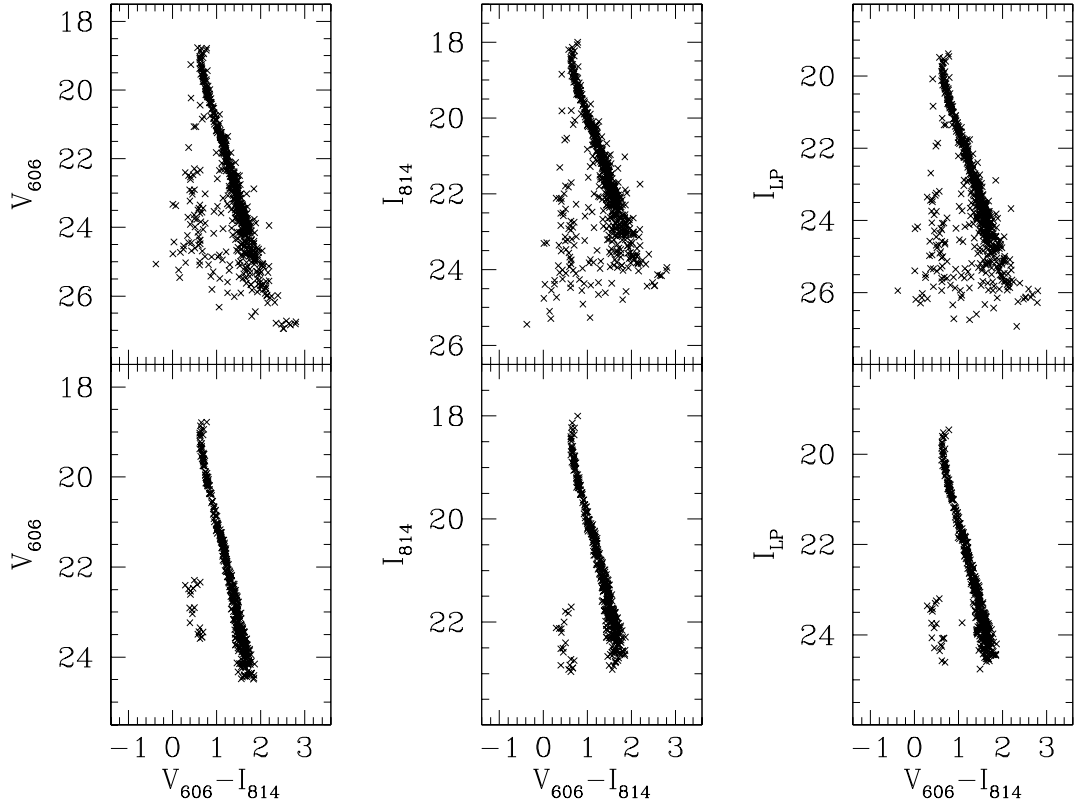,angle=270,height=4.5in,width=6in}
\caption{WFPC2/STIS color-magnitude diagrams (CMDs) of 47 Tuc and the
Small Magellanic Cloud (SMC).  The top panels show the CMDs for the stars
detected in both the F606W and F814W filters of WFPC2 and in
the LP filter of STIS.  The bottom panels show the subsets of stars from the
corresponding top panels which also lie on the WFPC2 main-sequence ridge lines
of 47~Tuc and the SMC (see Fig. 4).
Reddening and extinction corrections have not yet been applied to the
magnitudes and colors plotted here.}
\label{47tucallcmds}
\end{figure}

\begin{figure}[p]
\hspace*{0.25in}
\psfig{figure=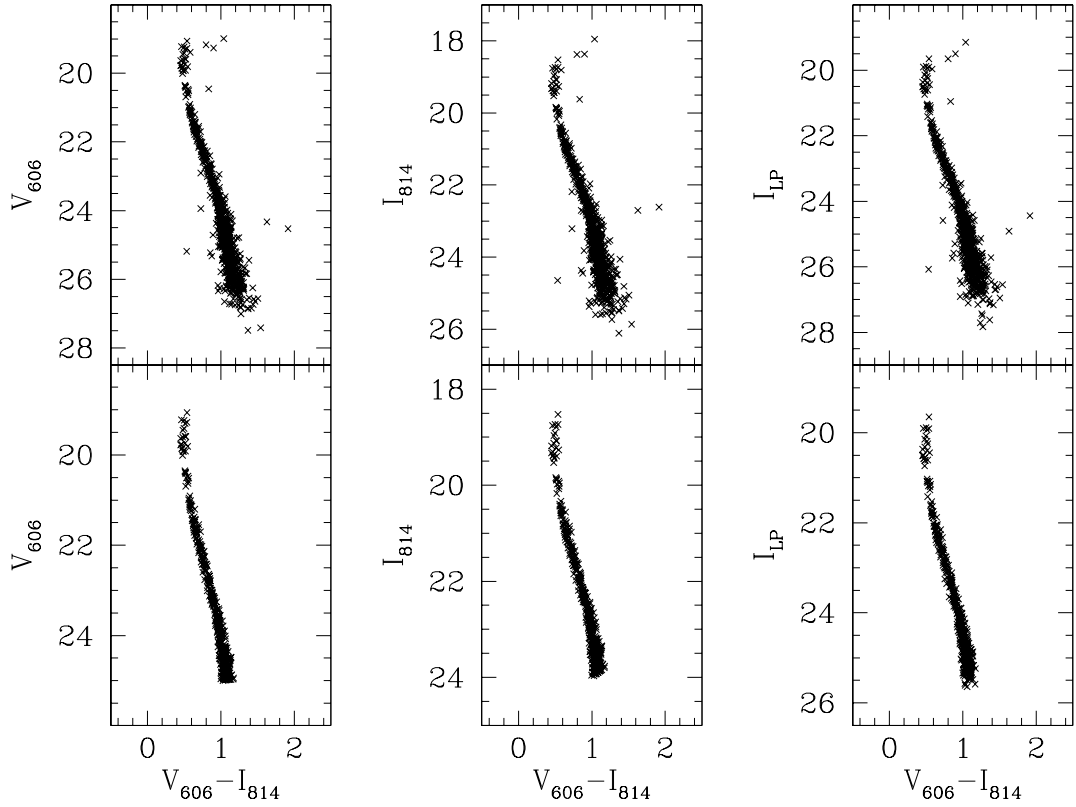,angle=270,height=4.5in,width=6in}
\caption{WFPC2/STIS color-magnitude diagrams (CMDs) of M15.  The top panels
show the CMDs for the stars detected in both the F606W and F814W filters of
WFPC2 and in
the LP filter of STIS.  The bottom panels show the subsets of stars from the
corresponding top panels which also lie on the WFPC2 main-sequence ridge lines
of M15 (see Fig. 5).
Reddening and extinction corrections have not yet been applied to the
magnitudes and colors plotted here.}
\label{m15allcmds}
\end{figure}

\begin{figure}[p]
\hspace*{0.25in}
\psfig{figure=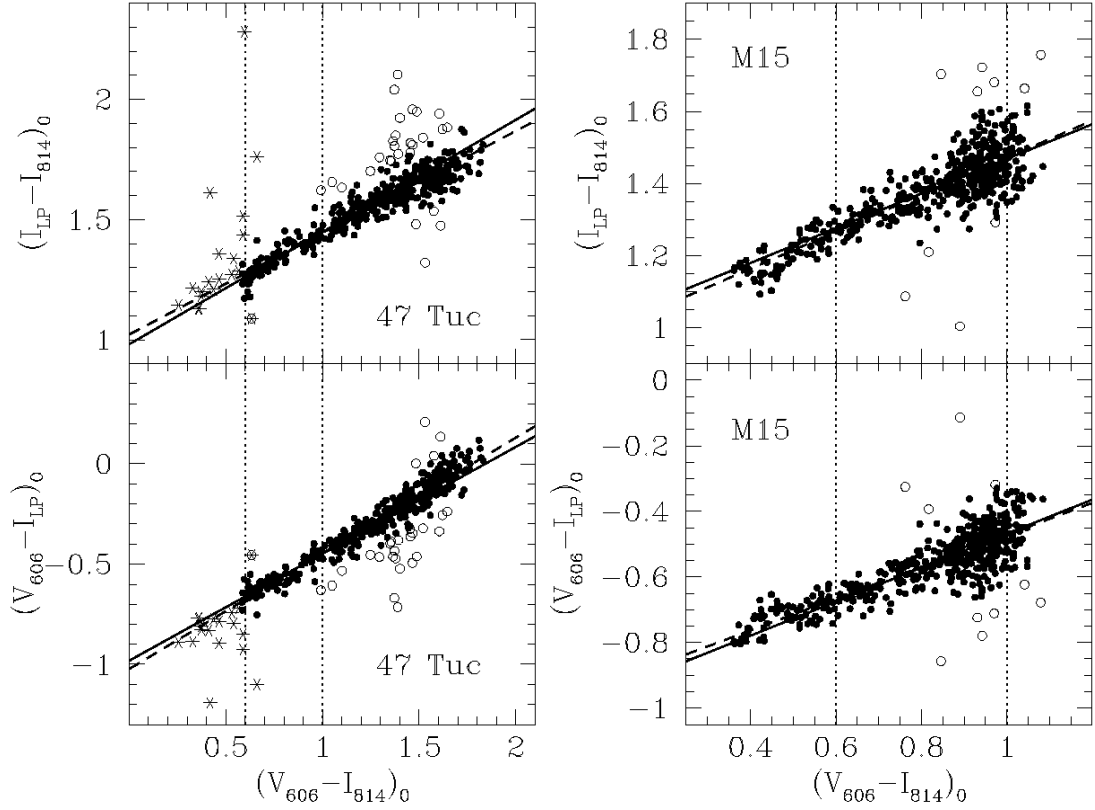,angle=270,height=4.5in,width=6in}
\caption{Color-color plots of 47 Tuc/SMC and M15.
The left-hand panels show the 47~Tuc and SMC stars
plotted in the bottom panels of Fig. 6 as circular points and asterisks,
respectively, and the right-hand panels show the M15 stars
plotted in the bottom panels of Fig. 7.  The dashed relations in the left-hand
and right-hand panels are the linear, least-squares fits to the 47~Tuc and M15
data, respectively, using iterative 3-$\sigma$ rejection;
the open points are those rejected during the fitting.
The solid lines are the analogous relations which result when only the
data between the vertical, dotted lines (0.6 $\leq$ (\mycolor)$_0$ $\leq$ 1.0)
are used in the fitting.  Coefficients of the fits are given in Table~3.}
\label{colorcolor}
\end{figure}

\begin{figure}[p]
\hspace*{0.75in}
\psfig{figure=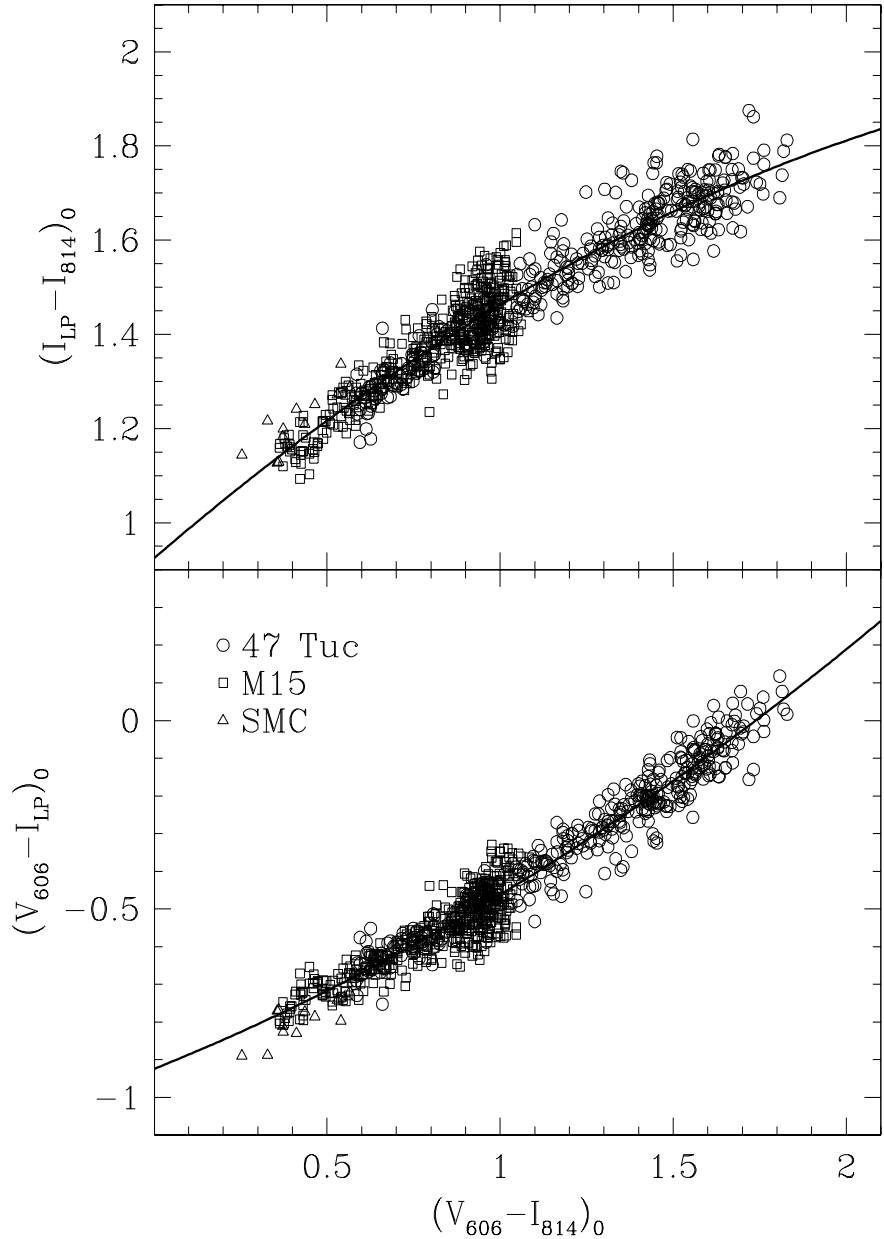,height=7in,width=5in}
\caption{The WFPC2-to-STIS transformations.  The main-sequence
ridge-line stars of 47~Tuc, M15 and the SMC are combined to derive
the transformations between the STIS \mystis\ magnitudes and the WFPC2
\myv\ and \myi\ magnitudes.  Stars from 47 Tuc are plotted as circles,
those from M15 are shown as squares, and SMC members are represented
by triangles.  The solid curves are the quadratic, least-squares fits
to the data, using iterative 3-$\sigma$ rejection; for clarity, points
rejected during the fitting are not shown.  The coefficients of the
transformation relations are given in Table~3.}
\label{finaltrans}
\end{figure}

\begin{figure}[p]
\hspace*{0.25in}
\psfig{figure=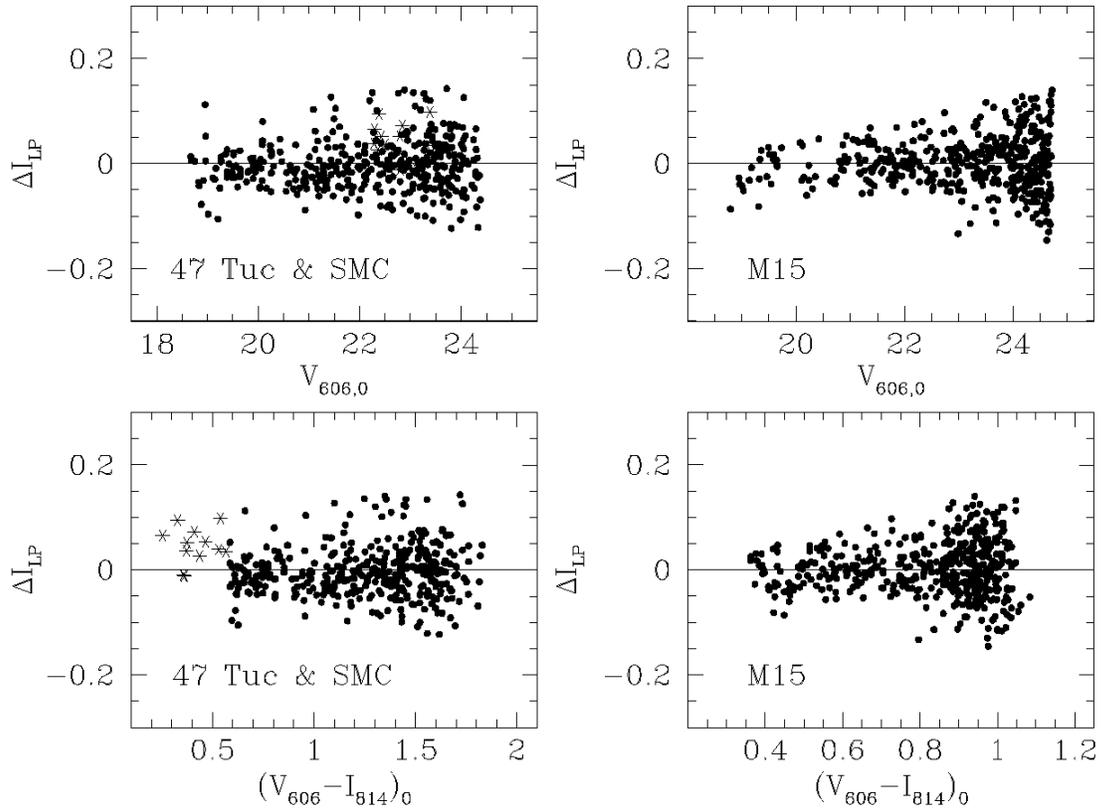,angle=270,height=4.5in,width=6in}
\caption{Differences between the predicted and observed I$_{\rm LP,0}$
magnitudes.  $\Delta$I$_{\rm LP}$ is the difference between the observed
I$_{\rm LP,0}$ magnitude of a star and that predicted from its WFPC2 V$_{606,0}$
magnitude and its (V$_{606}$--I$_{814}$)$_0$ color though the transformation
shown in the bottom panel of Fig. 9.  The left-hand panels show stars in
47~Tuc (filled points) and the SMC (asterisks), and the right-hand panels
show M15 stars.}
\label{magdevs}
\end{figure}

\clearpage

\begin{figure}[p]
\hspace*{0.75in}
\psfig{figure=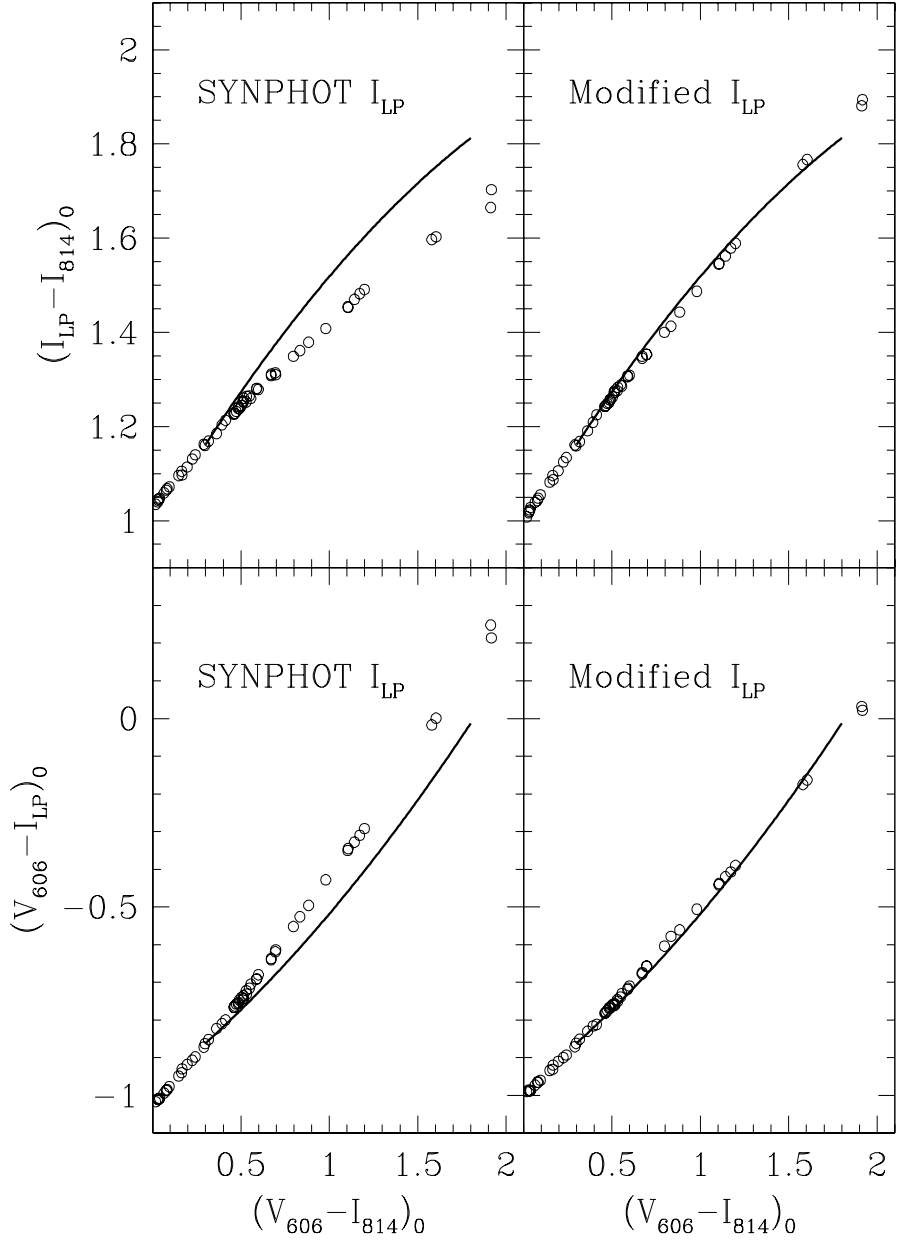,height=7in,width=5in}
\caption{Transformation relations from SYNPHOT.  The solid lines in all
panels are the final transformation relations derived here (see Figure~9 and
Table~3), truncated at the approximate color limits of the data used to derive
the relations; the open points are the calculated SYNPHOT colors of dwarf
stars in the BPGS Spectrophotometric Atlas.  The SYNPHOT data in the left-hand
panels has been calculated using the default SYNPHOT STIS longpass (LP) filter
response function (Leitherer et al. 2000), while the STIS LP filter response
suggested by Bessell (2000, private communication) was used to calculate the
SYNPHOT colors shown in the right-hand panels.  The SYNPHOT data and the
transformation relations have been normalized at (\mycolor)$_0$~=~0.3.}
\label{comp2syn}
\end{figure}

\begin{figure}[p]
\hspace*{0.25in}
\psfig{figure=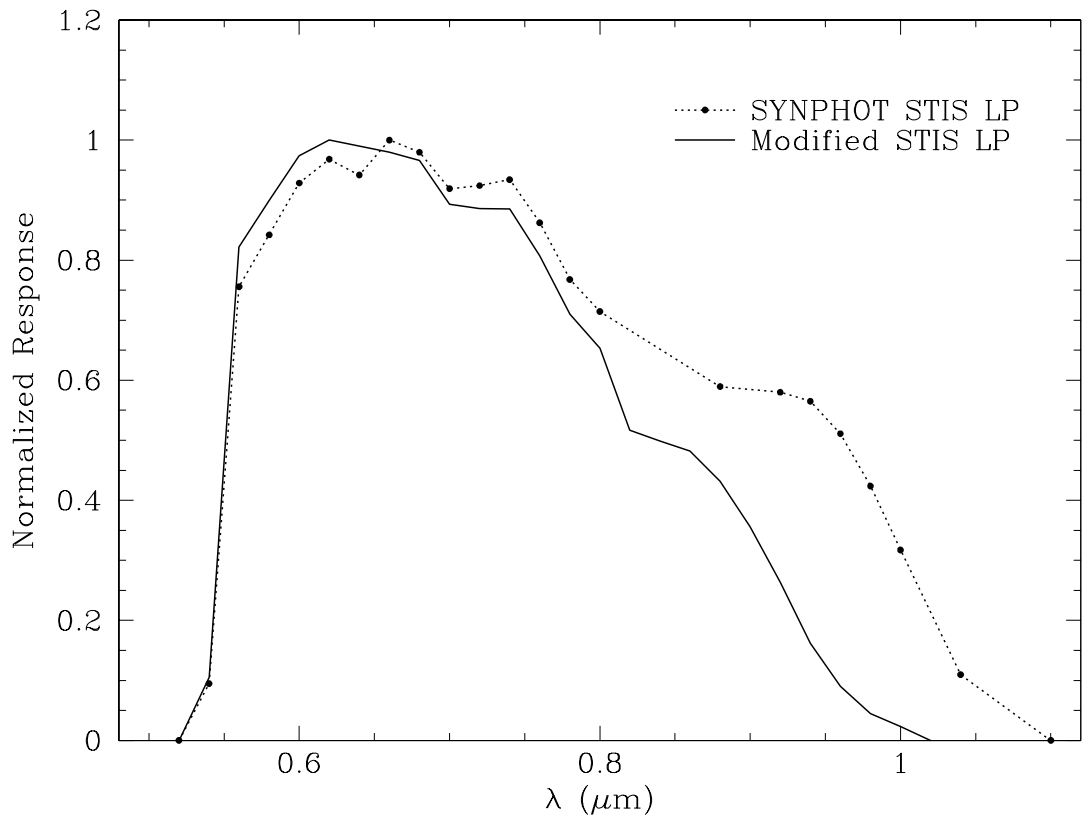,angle=270,height=4.5in,width=6in}
\caption{STIS LP filter response functions.  Response is defined as
$\lambda~\times$ throughput.  The dotted line is the normalized response of
the STIS longpass filter given in The STIS Instrument Handbook for Cycle 10
(Leitherer et al. 2000) and used in SYNPHOT; the solid line is the normalized
response of the modified STIS LP filter suggested by Bessell (2000, private
communication).}
\label{filters}
\end{figure}

\begin{figure}[p]
\hspace*{0.25in}
\psfig{figure=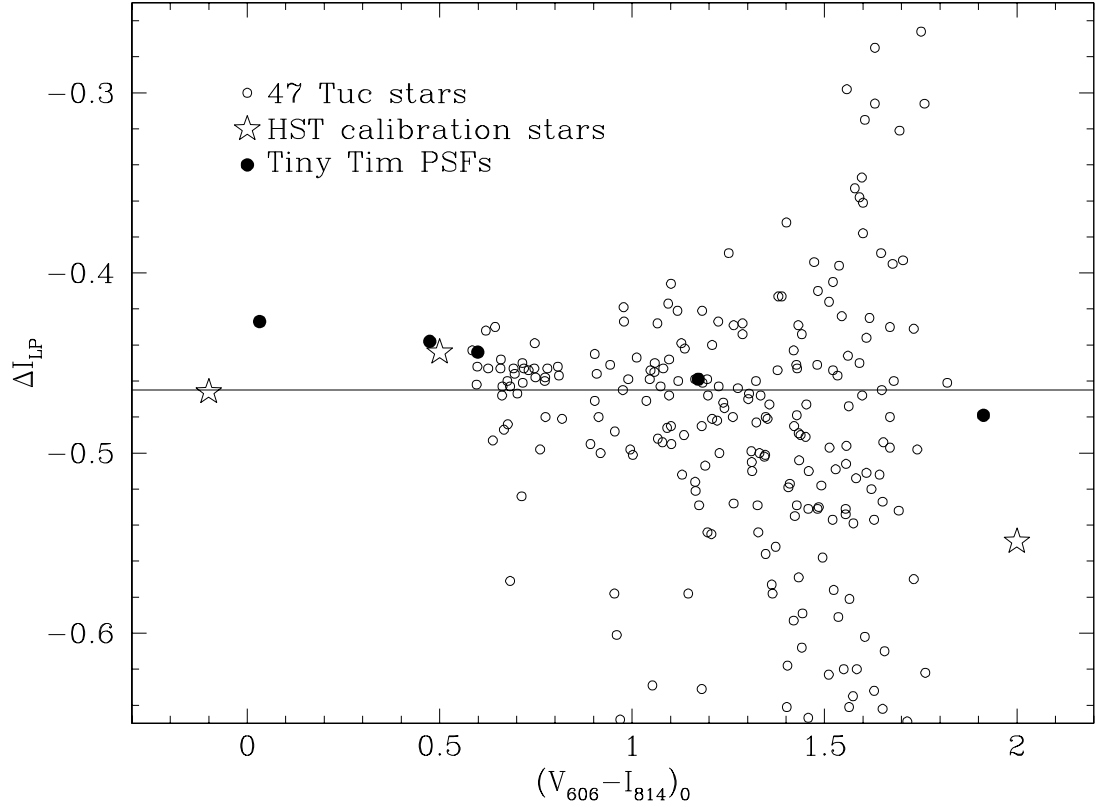,angle=270,height=4.5in,width=6in}
\caption{Aperture corrections as a function of color.  Aperture corrections
are shown for stars in 47~Tuc (open circles)
that were part of the sample used to derive our empirical transformations,
STIS calibration stars (filled circles), and Tiny Tim PSFs (asterisks).
From blue to red, the HST calibration stars are GRW~+70~5824, a white dwarf,
CPD~--60~7585, an F8 star, and BD~--11~3759, an M4 star, and the Tiny Tim PSFs
represent dwarf stars having spectral types A0, F8, G8, K7, and M3.
$\Delta$\mystis\ is the aperture correction required to transform a 2-pixel
(0.$^{\prime\prime}$1) magnitude into a magnitude measured in a
0.$^{\prime\prime}$5 aperture; the
horizontal line shows the corresponding aperture correction applied to
our 2-pixel magnitudes of 47~Tuc stars (see Table~2).}
\label{apcorrs}
\end{figure}

\end{document}